%% file: usenixsecurity2026.tex
\documentclass[letterpaper,twocolumn,10pt]{article}
\usepackage{usenix}

\usepackage{tikz}
\usepackage{amsmath}

\usepackage{filecontents}

\usepackage{amsthm}
\usepackage{amsmath}
\usepackage{amssymb}
\usepackage{mathrsfs}  
\usepackage{color}
\usepackage{float}  
\usepackage{makecell}
\usepackage{multirow}
\usepackage{enumitem}

\usepackage{subfigure}  
\usepackage{hyperref }
\usepackage{url}
\usepackage{bm}

\usepackage{booktabs}
\usepackage{footmisc}
\usepackage[ruled]{algorithm2e}

\usepackage{tcolorbox}

\newtheorem{theorem}{Theorem}
\newtheorem{lemma}{Lemma}

\begin{document}

\date{}

\title{\Large \bf
Efficient and High-Accuracy Secure Two-Party Protocols for a Class of Functions with Real-number Inputs
}

    \author{
{\rm Hao Guo$^{1\dagger}$, Zhaoqian Liu$^{1\dagger}$, Liqiang Peng$^{2}$\textsuperscript{*}, Shuaishuai Li$^{3}$, Ximing Fu$^{4}$\textsuperscript{*}, Weiran Liu$^{2}$, Lin Qu$^{2}$}\\[0.5ex]
    \rm $^{1}$The Chinese University of Hong Kong, Shenzhen\\
    \rm $^{2}$Alibaba Group\\
    \rm $^{3}$Zhongguancun Laboratory, Beijing, China\\
    \rm $^{4}$Key Laboratory of Cyberspace and Data Security, Ministry of Emergency Management\\
    \rm guohao.g@outlook.com, zhaoqianliu@link.cuhk.edu.cn, 
    fuximing@hit.edu.cn
    } 

\maketitle

\begin{abstract}



In two-party secret sharing scheme, values are typically encoded as unsigned integers $\mathsf{uint}(x)$, whereas real-world applications often require computations on signed real numbers $\mathsf{Real}(x)$. 
To enable secure evaluation of practical functions, it is essential to computing $\mathsf{Real}(x)$ from shared inputs, as protocols take shares as input.
At USENIX'25, Guo et al. proposed an efficient method for computing signed integer values $\mathsf{int}(x)$ from shares, which can be extended to compute $\mathsf{Real}(x)$.
However, their approach imposes a restrictive input constraint $|x| < \frac{L}{3}$ for $x \in \mathbb{Z}_L$, limiting its applicability in real-world scenarios.
In this work, we significantly relax this constraint to $|x| < B$ for any $B \leq \frac{L}{2}$, where $B = \frac{L}{2}$ corresponding to the natural representable range in $x \in \mathbb{Z}_L$.
This relaxes the restrictions and enables the computation of $\mathsf{Real}(x)$ with loose or no input constraints.
Building upon this foundation, we present a generalized framework for designing secure protocols for a broad class of functions, including integer division ($\lfloor \frac{x}{d} \rfloor$), trigonometric ($\sin(x)$) and exponential ($e^{-x}$) functions.
Our experimental evaluation demonstrates that the proposed protocols achieve both high efficiency and high accuracy.
Notably, our protocol for evaluating $e^{-x}$ reduces communication costs to approximately 31\% of those in SirNN (S\&P’21) and Bolt (S\&P’24), with runtime speedups of up to $5.53 \times$ and $3.09 \times$, respectively.
In terms of accuracy, our protocol achieves a maximum ULP error of $1.435$, compared to $2.64$ for SirNN and $8.681$ for Bolt.

\end{abstract}

\renewcommand{\thefootnote}{\fnsymbol{footnote}} 

\input{sec/Intro_NEW}

\input{sec/Pre_NEW}

\input{sec/MW}

\input{sec/APP}

\input{sec/experiment_NEW}

\input{sec/Conclusion}


\cleardoublepage

\appendix
\section*{Ethical Considerations}

All authors of this paper unanimously agree and declare the following:

\begin{itemize}
    \item 
    We attest that we have read the ethics discussions in the conference call for papers, the detailed submissions instructions, and the ethics guidelines.

    \item 
    We affirm that our research adheres to all applicable ethical standards and the principles of the Open Science Policy. 
    This study does not involve human participants, data privacy issues, or other sensitive matters; consequently, the requirement for informed consent does not apply.
    Additionally, we confirm that the planned future work by our team (e.g., post-publication activities) complies with ethical principles.
    
\end{itemize}



\section*{Open Science}


We fully endorse the principles of Open Science Policy. 
To promote transparency and reproducibility, we have integrated our research artifacts into an open-source repository on GitHub, which is now ready for public access. 
In the final version of our paper, we commit to openly sharing all research materials, granting the scientific community unrestricted access to review, validate, and build upon our work





\bibliographystyle{plain}
\bibliography{bib.bib}

\appendix

\input{App/proofs}

\end{document}

%% file: sec/Intro_NEW.tex
\section{Introduction}

Secure two-party computation (2PC) \cite{2008How,YAO82,MPC3 } is a foundational cryptographic primitive that enables two mutually distrustful parties to collaboratively evaluate a public function over their private inputs without revealing any information beyond the function's  output.
By guaranteeing both correctness and privacy, 2PC has become a foundational technology for privacy-preserving computation, enabling a wide range of real-world applications such as secure data analysis and privacy-preserving machine learning.
However, the practical deployment of 2PC is often limited by significant communication overhead, largely driven by expensive cryptographic operations such as those based on public-key primitives. Consequently, the design of efficient and scalable two-party protocols has become a central focus in modern 2PC research.

The Garbled Circuits (GC) method, introduced by Yao \cite{YAO82}, was the first general solution proposed for 2PC. 
While GC provides a robust framework, its high computational and communication costs limit its practicality in real-world scenarios.
Recent advances in privacy-preserving machine learning (PPML) \cite{SecureML,ABY3,MiniONN,CipherGPT,bolt,Falcon} have spurred the development of customized protocols that prioritize efficiency. 
Additionally, innovative protocol design principles have emerged to guide the construction of more effective solutions. 
For instance, SirNN \cite{SiRnn} leverages the concept of non-uniform bitwidths to significantly reduce the overhead of various protocols and secure RNN inference tasks. 
The fitting techniques, such as polynomial fitting and Fourier series fitting, have become standard tools for evaluating non-linear functions, including $\mathsf{GELU}$ \cite{iron, bolt}, $e^{-x}$ \cite{SiRnn} and $\mathsf{sigmoid}$ \cite{sin_SHAFT, Squirrel}.
The recent work SEAF \cite{SEAF} further advances this area by providing an optimized protocol design method for general non-linear activation functions.

In the two-party secret sharing scheme, an input is encoded as unsigned fixed-point number $x = \mathsf{uint}(x)$ to perform cryptographic operations, and shared as $x = x_0 + x_1 \bmod L$.
However, most real-world functions operate on signed real numbers $\mathsf{Real}(x)$.
Consequently, a conversion from $\mathsf{uint}(x)$ to $\mathsf{Real}(x)$ is necessary.
Since protocols have access only to the shares $x_0$ and $x_1$, $\mathsf{Real}(x)$ need to be computed from these shares.
Guo et al. \cite{geo} introduced a new variable $\mathsf{MW}(x)$ to derive the signed value $\mathsf{int}(x)$ from shares via $\mathsf{int}(x) = x_0 + x_1 - \mathsf{MW}(x) \cdot L$.
The corresponding signed real value is then  $\mathsf{Real}(x) = \frac{\mathsf{int(x)}}{2^f}$, where $f$ denotes the number of fractional bits.
However, the approach of Guo et al. has two main limitations.
First, they only give an efficient method for computing $\mathsf{MW}(x)$ under the constraint  $|x| < \frac{L}{3}$, which is often violated in practice and thus limits applicability.
Second, they focus on computing signed integer values, whereas many real-world applications require real number inputs.



In this work, we first generalize the constraint for computing $\mathsf{MW}(x)$ from $|x| < \frac{L}{3}$ to $|x| < B$ for $B \leq \frac{L}{2}$, and then extend this method to compute $\mathsf{Real}(x)$.
When an upper bound $B$ on $|x|$ is known a priori,  this relaxed constraint expands the range of applicable values.
In the absence of such prior knowledge, we set $B = \frac{L}{2}$, which is the natural upper bound of $|x|$ for $x \in \mathbb{Z}_{L}$.
Once $\mathsf{MW}(x)$ is obtained, the signed real number with $f$-bit precision can be written as $\mathsf{Real}(x) = \frac{x_0 + x_1 - \mathsf{MW}(x)\cdot L}{2^f}$.
Then a function $\mathsf{func}(\cdot)$ with $\mathsf{Real}(x)$ as input can be written as $\mathsf{func}(\mathsf{Real}(x)) = \mathsf{func}(\frac{x_0}{2^f} + \frac{x_1}{2^f} + \frac{-\mathsf{MW}(x) \cdot L}{2^f})$.
Based on this representation, we present a general protocol design method for a broader class of practical functions, including trigonometric and exponential operations.
Our main contributions are summarized as follows:
\begin{itemize}


    \item 
    We first optimize the comparison protocol for the scenarios with constraints on the input range.
    Based on this, we relax the constraint for computing $\mathsf{MW}(x)$ from $|x| < \frac{L}{3}$ in prior work to $|x| < B$, where $B \leq \frac{L}{2}$.
    Notably, when $B =\frac{L}{2}$, the only constraint on $x$ is that $x \in \mathbb{Z}_{L}$.
    Then the real number $\mathsf{Real}(x)$ can be computed from shares efficiently.

    \item 
    We propose a novel method for securely evaluating a class of functions $\mathsf{func}(\cdot)$ which take $\mathsf{Real}(x)$ as input and satisfy the property $\mathsf{func}(a + b + c) = \sum_i f_i(a) \cdot g_i(b) \cdot h_i(c)$.
    We instantiate this method to several important  real-world functions, including integer division ($\lfloor \frac{x}{d} \rfloor$), trigonometric ($\sin(x)$) and exponential ($e^{-x}$) functions.
    The implementations demonstrate that our new protocols achieve both low overhead and high accuracy.

\end{itemize}

\subsection{Our results}

\vspace{5pt}
\noindent
\textbf{Comparison protocol with constraint.}
A comparison protocol takes $l$-bit $x$ from party $P_0$ and $l$-bit $y$ from party $P_1$ as inputs, and outputs $b = \bm{1}\{x < y\}$, with linear communication complexity $O(l)$ \cite{CrypTFlow2}.
In this work, we focus on comparison protocols with constraints.
Specifically, for $x, y \in \mathbb{Z}_{L}$, where $L = 2^l$, we assume that $x$ is either smaller than $y$ or greater than $y$ by $A$.
Formally, $x - y \in [A, L) \bigcup [-L,0)$.
Under this constraint, we prove that $y < x$ if and only if $\lfloor \frac{y}{A} \rfloor < \lfloor \frac{x}{A} \rfloor$.
Consequently, the input length of the comparison protocol is reduced from $l$ to $\lceil \log \lfloor \frac{L}{A} \rfloor \rceil$, and the communication complexity decreases from $O(l)$ (or $O(\lceil \log L \rceil)$) to $O(\lceil \log \lfloor \frac{L}{A} \rfloor \rceil)$.

\vspace{5pt}
\noindent
\textbf{Computing $\mathsf{MW}(x)$ for $|x| < \frac{L}{2}$.}
We generalize Guo et al.'s work \cite{geo} for computing $\mathsf{MW}(x)$ from constraint $|x| <\frac{L}{3}$ to $|x| < B$ for any $B \leq \frac{L}{2}$.
$B = \frac{L}{2}$ implies no additional constraints beyond $x \in \mathbb{Z}_L$.
The theoretical communication of our protocol for computing $\mathsf{MW}(x)$ is summarized in Table~\ref{tab:comm MW}, and some instances are listed in Table~\ref{tab:comm MW 222}.
When $B = 0.9999 \cdot \frac{L}{2}$ (allowing $x$ to take $99.99\%$ of the values in $\mathbb{Z}_{L}$), only a 
$14$-bit input comparison protocol is required.
Even for $B = 0.999999 \cdot \frac{L}{2}$ (covering $99.9999\%$ of values), the input length is $20$ bits, achieving $\frac{l}{l*} = \frac{37}{20} =1.85$ times improvement compared to the method with natural constraint $B = \frac{L}{2}$, where $l$-bit comparison protocol is invoked.

Moreover, for the case an $l$-bit $x$ is shared over larger ring $\mathbb{Z}_{2^{l_r}}$, where $l_r \geq l + 1$, we give an efficient method to compute $\mathsf{MW}(x \bmod 2^{l}, 2^l)$ from $\mathsf{MW}(x, 2^{l_r})$, where the latter can be easily computed.
Then, computations can be performed on small ring $\mathbb{Z}_{2^{l}}$ with low overhead.
This technique significantly improves the efficiency for evaluating $e^{-x}$ in subsection~\ref{subsec:e^-x}.


\begin{table}[htb]
    \centering
    \caption{Theoretical communication for our computing $\mathsf{MW}(x)$ protocol $\Pi_{\mathsf{MW}}^{l,l'}$ with constraint $|x| < B$.
    $\Pi_{\mathsf{MW}}^{l,l'}$ takes shared $l$-bit $x$ as input and outputs shared $l'$-bit $\mathsf{MW}(x)$.
    Parameters are defined as $L=2^l$, $K = \lfloor \frac{L}{L-2B} \rfloor$.
    When $B \neq \frac{L}{2}$, let $l^* = \lceil \log \lfloor \frac{L}{L-2B} \rfloor \rceil$; when $B = \frac{L}{2}$, let $l^* = l$.
    }
    \label{tab:comm MW}
    \subtable[Theoretical communication for $x\in\mathbb{Z}_{L}$.]{
\setlength{\tabcolsep}{24pt}
    \begin{tabular}{cc}
        \toprule
        Range of $B$ & Comm. \\
        \midrule
        $[0, \frac{3}{8} L)$ & $K \cdot (\lambda + l')$ \\
        $[\frac{3}{8} L, \frac{L}{2}]$ & $<\lambda (l^* + 1) + 14 l^* + l'$\\
        \bottomrule
    \end{tabular}
    \label{tab:comm MW 111}
    }

    \qquad

    \subtable[Theoretical communication for some examples, where $l=37$.]{
\setlength{\tabcolsep}{24pt}
    \begin{tabular}{ccc}
        \toprule
        Value of $B$ & $l^*$ & Comm. \\
        \midrule
        $0.5 \cdot \frac{L}{2}$ & - & $165$ \\
        $0.8 \cdot \frac{L}{2}$ & 3 & $<591$ \\
        $0.9999 \cdot \frac{L}{2}$ & 14 & $<2153$ \\
        $0.999999 \cdot \frac{L}{2}$ & 20 & $<3005$ \\
        $1 \cdot \frac{L}{2}$ & 37 & $<5254$ \\
        \bottomrule
    \end{tabular}
    \label{tab:comm MW 222}
    }
\end{table}

\begin{table}[htb]
    \centering
    \caption{
    Theoretical communication for our division, exponential, and trigonometric protocols.
    The input bitwidth is $l$, with precision $f$.
    $C_{\mathsf{MW}}$ denotes the communication of computing $\mathsf{MW}(x)$ protocol.
    $l_d = \lceil \log d \rceil$,  where $d$ is the divisor.
    $\Pi_{\mathsf{rExp}}^1$ denotes that the protocol takes input with range $\mathsf{Real}(x) \in [0,8)$, and the protocol marked by $\Pi_{\mathsf{rExp}}^2$ works for any $\mathsf{Real}(x) \geq 0$.
    }
    \label{tab:comm app}
\setlength{\tabcolsep}{8pt}
    \begin{tabular}{ccc}
        \toprule
        Func. & Protocol & Comm. \\
        \midrule
        
        \multirow{1}*{$\lfloor \frac{x}{d} \rfloor$} 
        & $\Pi_{\mathsf{Div}}$, Sec~\ref{sec:app div} & $\lambda (l_d + 3) + 5l + 18l_d + C_{\mathsf{MW}}$ \\

        \hline


        \multirow{1}*{$\sin(x)$} 
        & $\Pi_{\mathsf{sin}}$, Sec~\ref{sec:app sin} & $69 \lambda + 1888 + C_{\mathsf{MW}}$ \\

        \hline

        \multirow{2}*{$e^{-x}$} 
        & $\Pi_{\mathsf{rExp}}^1$, Sec~\ref{sec:app exp} & $28 \lambda + 2l + 4f + 897$ \\
        & $\Pi_{\mathsf{rExp}}^2$ Sec~\ref{sec:app exp} & $\lambda (l + 29) + 18l + 4f  + 897$ \\

        \bottomrule
    \end{tabular}
\end{table}

\vspace{5pt}
\noindent
\textbf{New protocol designs for real-world functions.}
Using $\mathsf{MW}(x)$ as a foundation, we can compute the signed real number from shares.
Further, based on the representation of $\mathsf{Real}(x)$, we propose a method for evaluating functions $\mathsf{func}(\cdot)$ with the property $\mathsf{func}(a+b+c) = \sum_{i=0}^{k-1} f_i(a) \cdot g_i(b) \cdot h_i(c)$.
This method is applied to several real-world functions, including integer division ($\lfloor \frac{x}{d} \rfloor$), trigonometric ($\sin(x)$) and exponential functions ($a^x$, $e^{-x}$).
The theoretical communication for these protocols is detailed in Table~\ref{tab:comm app}.
The experimental results in Section~\ref{sec:exp app} demonstrate significant improvements.
For our division protocol, we achieve $1.4 \times$ to $4.84 \times$ improvement over CrypTFlow2 \cite{CrypTFlow2}.
For the exponential function protocol, compared to SirNN \cite{SiRnn} and Bolt \cite{bolt}, we reduce communication costs to $31\%$, while improving runtime by up to $5.53 \times$ and $3.09 \times$, respectively.
Furthermore, when applied on the activation function evaluation in PPML, our $\mathsf{Softmax}$ protocol, built on the optimized $e^{-x}$ and batch division protocols, achieves $4.59 \times$ to $6.2 \times$ improvement over Iron \cite{iron}, and $1.94 \times$ to $2.32 \times$ improvement over Bolt \cite{bolt}.

Our protocols not only reduce overhead but also achieve higher accuracy.
Specifically, the maximum ULP (units in the last place) error of our $\sin(x)$ protocol is approximately $1.3$.
Our $e^{-x}$ protocol achieves a maximum ULP error of $1.435$, outperforming prior works, which reports ULP errors of $2.64$ in SirNN and $8.681$ in Bolt.

\section{Overview}


\subsection{Computing $\mathsf{Real}(x)$ from Shares}

Guo et al. \cite{geo} introduced $\mathsf{MW}(x)$ to compute the signed value $\mathsf{int}(x)$ from shares as $\mathsf{int}(x) = x_0 + x_1 - \mathsf{MW}(x) \cdot L$.
The $\mathsf{MW}(x)$ is an abbreviation of $\mathsf{MW}(x_0, x_1, L)$ or $\mathsf{MW}(x, L)$, with the constraint $B \leq \frac{L}{2}$, $\mathsf{MW}(x)$ can be computed as:
\begin{equation}\label{equ:MW def1}
\begin{split}
\mathsf{MW}(x) =& \mathsf{MW}(x,L) = \mathsf{MW}(x_0, x_1,L)  \\
    =&\begin{cases}
        0, \quad \text{if}~ x_0+x_1 \in [0, B),\\
        1, \quad \text{if}~  x_0+x_1 \in [L-B, L+B),\\
        2, \quad \text{if}~  x_0+x_1 \in [2L-B, 2L).\\
    \end{cases}
\end{split}
\end{equation}
Using $\mathsf{MW}(x)$, the real value can be computed as 
\begin{equation}\label{equ:real}
    \mathsf{Real}(x) = \frac{\mathsf{int}(x)}{2^f} = \frac{x_0 + x_1 - \mathsf{MW}(x) \cdot L}{2^f},
\end{equation}
where $f$ denotes the precision.
Therefore, the core for computing $\mathsf{Real}(x)$ is to compute $\mathsf{MW}(x)$.
Guo et al. \cite{geo} proposed an efficient method to compute $\mathsf{MW}(x)$ under constraint $|x| < \frac{L}{3}$.
However, many practical scenarios do not allow this constraint to be satisfied.

Our work focuses on relaxing the constraint to $|x| < B$ for any $B \leq \frac{L}{2}$, and therefore computing $\mathsf{Real}(x)$ with any constraint.
In this case, determining whether $\mathsf{MW}(x) = 0$ is equivalent to checking whether $x_0 + x_1 < B$.
Let $\alpha = x_0$ and $\beta = B - x_1$, we have that $\mathsf{MW}(x) = 0$ if and only if $\alpha < \beta$, which can be determined using a comparison protocol.
Moreover, there is a constraint on the input of this comparison protocol, as $\alpha - \beta = x_0 + x_1 \in [-B,0) \bigcup [L - 2B) \bigcup [2L - 2B, 2L-B)$.
Therefore, determining $\mathsf{MW}(x) = 0$ reduces to performing a comparison protocol with this specialized constraint. 
In this work, we propose new protocols for comparison protocol under such constraint in Section~\ref{sec:comp}, and demonstrate how to compute $\mathsf{MW}(x)$ by invoking only one comparison protocol in Section~\ref{sec: subsection MW}.
Finally, $\mathsf{MW}(x)$ can be efficiently computed even if $B$ is very close to $\frac{L}{2}$.
For the case $B = \frac{L}{2}$, $\mathsf{MW}(x)$ is computed by invoking only one $l$-bit comparison protocol.

Furthermore, we consider the scenario in which an $l$-bit $x$ is shared over the ring $\mathbb{Z}_{2^{l_r}}$, where $l_r \geq l + 1$.
In this case, $\mathsf{MW}(x, 2^{l_r})$ can be computed efficiently since $|x| < \frac{2^{l_r}}{4}$ \cite{geo}.
However, protocols are more efficient when executed over the small ring $\mathbb{Z}_{2^l}$ instead of $\mathbb{Z}_{2^{l_r}}$, due to the shorter input lengths involved. 
To benefit from the advantages of both approaches, we propose a method to derive $\mathsf{MW}(x \bmod 2^{l}, 2^{l})$ from $\mathsf{MW}(x, 2^{l_r})$, as detailed in Section~\ref{sec: subsection MW ring conversion}.
Using this method, we first compute $\mathsf{MW}(x, 2^{l_r})$ over the ring $\mathbb{Z}_{2^{l_r}}$, then convert it to $\mathsf{MW}(x \bmod 2^{l}, 2^{l})$, and finally perform protocols over $\mathbb{Z}_{2^{l}}$.
This approach allows us to achieve the high efficiency for computing $\mathsf{MW}(x, 2^{l_r})$ and benefits of performing protocols with shorter input lengths.




\subsection{Applications on Real-world Functions}
\label{subsec:app}

Consider a class of functions $\mathsf{func}(\cdot)$ with the following properties: 
$\mathsf{func}(\cdot)$ takes a single real number $x$ as input, by dividing $x$ as $x = a+b+c$, $\mathsf{func}(x)$ can be computed as:
\begin{equation}\label{equ:func x y z}
    \mathsf{func}(x) = \mathsf{func}(a+b+c) = \sum_{i=0}^{k-1} f_i(a) \cdot g_i(b) \cdot h_i(c), 
\end{equation}
where $f_i$, $g_i$ and $h_i$ are public functions.
For example, exponential function is an instance of $\mathsf{func}(\cdot)$, as $e^{a+b+c} = e^a \cdot e^b \cdot e^c$.
In this case, $k = 1$, and $f_0$, $g_0$ and $h_0$ are all exponential functions.
Using the real number representation in Equation~\ref{equ:real}, $\mathsf{func}(\cdot)$ with the form of Equation~\ref{equ:func x y z} can be computed as: 
\begin{equation}\label{equ:func int}
\mathsf{func}(\mathsf{Real}(x)) =  \sum_{i=0}^{k-1} f_i(\frac{x_0}{2^f}) \cdot g_i(\frac{x_1}{2^f}) \cdot h_i(\frac{-\mathsf{MW}(x) \cdot L}{2^f}).
\end{equation}
In this formulation, $f_i(\frac{x_0}{2^f})$ and $g_i(\frac{x_1}{2^f})$ can be computed locally by parties $P_0$ and $P_1$, respectively.
$h_i(\frac{-\mathsf{MW}(x) \cdot L}{2^f})$ depends on $\mathsf{MW}(x)$, which is a shared value.
However, since $\mathsf{MW}(x) \in \{0,1,2\}$, $h_i(\frac{-\mathsf{MW}(x) \cdot L}{2^f})$ can take only three possible values.
To compute Equation~\ref{equ:func int} efficiently, we let parties to compute $w_i^j=f_i(\frac{x_0}{2^f}) \cdot g_i(\frac{x_1}{2^f}) \cdot h_i(\frac{-j \cdot L}{2^f})$ for $j \in \{0,1,2\}$.
Here, $h_i(\frac{-j \cdot L}{2^f})$ is public and known to both parties, and therefore multiplying $h_i(\frac{-j \cdot L}{2^f})$ can be computed very efficiently.
We store the value $w_i^j$ in a lookup table $T$, and use $\mathsf{MW}(x)$ as the index to retrieve $w_i=f_i(\frac{x_0}{2^f}) \cdot g_i(\frac{x_1}{2^f}) \cdot h_i(\frac{-\mathsf{MW}(x)\cdot L}{2^f})$.
Finally, by summing over all $w_i$ for $i \in \{0,...,k-1\}$, we can get the result of $\mathsf{func}(\mathsf{Real}(x))$.
The main overhead of this method involves several multiplications, one computation of $\mathsf{MW}(x)$, and one lookup table protocol.
Moreover, the bitwidths and precisions of the components $f_i(\frac{x_0}{2^f})$, $g_i(\frac{x_1}{2^f})$ and $h_i(\frac{-\mathsf{MW}(x) \cdot L}{2^f})$ are independent of the input bitwidth and precision, and can be set arbitrarily.
Therefore, highly accurate approximations can be derived by setting high-precision of these components.

If a function has the property of Equation~\ref{equ:func x y z}, it can be securely evaluated as Equation~\ref{equ:func int}.
Examples include truncation, division, trigonometric functions, and exponential functions.
Thus, these functions can all be evaluated using this method.
The specific protocol designs are listed in Section~\ref{sec:app all}.

%% file: sec/Pre_NEW.tex
\section{Preliminaries}

\textbf{Notations.}
This work considers power-of-two ring $\mathbb{Z}_L$, where $L = 2^l$.
The indicator function $\bm{1}\{\text{condition}\}$ returns $1$ if the condition holds and $0$ otherwise.
For an unsigned value $x = \mathsf{uint}(x) \in \mathbb{Z}_L$, its signed representation is $\mathsf{int}(x) = \mathsf{uint}(x) - \mathsf{MSB}(x) \cdot L$, where $\mathsf{MSB}(x)$ denotes the most significant bit (MSB).
We use $[\![ \cdot ]\!]$ to denote the two-party additive secret sharing.
Specifically, $[\![ \cdot ]\!]^l$ denotes arithmetic sharing over $\mathbb{Z}_L$, and $[\![ \cdot ]\!]^B$ denotes Boolean sharing.
For $x \in \mathbb{Z}_L$, we define the constraint $|x| < B$ as $x \in [0, B) \cup [L - B, L)$.
Note that this differs from the traditional absolute value definition: in our setting, we allow $x$ to take value of $L - B$.
Let $S$ as a set, $A$ is a subset of $S$, and symbol $S \setminus A$ denotes the difference set $\{x | x \in S, x \notin A\}$.
$x \gg k$ denotes the right shift $k$-bit operation.
The security parameter $\lambda$ is set to $128$ throughout this work.

\vspace{5pt}
\noindent
\textbf{Fixed-point representation.}
To perform secure computation over real numbers, values are first encoded as integers over a ring or field using fixed-point representation.
Given a real number ${x_{real}} \in \mathbb{R}$ represented with $l$ bits, and a fractional precision of $f$ bits, it is encoded as $x = \mathsf{Fix}(x_{real},l,f) = \lfloor x_{real} \cdot 2^f \rfloor \bmod 2^l$.
To recover the original signed real value $x_{real}$ from its unsigned fixed-point representation $x$, the signed value $\mathsf{int}(x) = x - \mathsf{MSB}(x) \cdot L$ is first computed, then $x_{real} = \mathsf{Real}(x, l, f) = \frac{\mathsf{int}(x)}{2^f}$.

\vspace{5pt}
\noindent
\textbf{ULP error.}
Suppose $a$ is a real number and $\widetilde{a}$ is an approximation of $a$ with precision $\widetilde{f}$-bit.
This work uses the "units in last place" (ULP) error \cite{ULP} to measure the approximation error between $a$ and $\widetilde{a}$, which quantifies the number of minimum representable units between the two values.
Let the minimum representable unit of $a$ and $\widetilde{a}$ as $u$ and $\widetilde{u}$, respectively.
They can be expressed as  $a = k \cdot u$ and $\widetilde{k} \cdot \widetilde{u}$, where both $k$ and $\widetilde{k}$ are integers.
The ULP error is then computed as $\frac{|a - \widetilde{a}|}{\widetilde{u}}$.
Previous works \cite{SiRnn, bolt} typically set $u = \widetilde{u} = 2^{-\widetilde{f}}$, resulting in integer-valued ULP errors.
In contrast, we set $u = 10^{-6}$ while maintaining $\widetilde{u} = 2^{-\widetilde{f}}$, which yields fractional ULP errors and allows for more precise error measurement.




\subsection{Two-party Secure Computation}

\subsubsection{Two-party additive secret sharing}

In two-party additive secret sharing, a secret value $x$ is shared as $x = x_0 + x_1 \bmod L$, where participant $P_0$ holds $x_0$ and $P_1$ holds $x_1$.
This sharing can be represented as $[\![ x ]\!]^{l} = ([\![ x ]\!]_0^{l}, [\![ x ]\!]_1^{l})$, where $[\![ x ]\!]_i^{l}$ is usually abbreviated as $x_i$ for $i \in \{0,1\}$.
The state "$P_0$ and $P_1$ hold $[\![ x ]\!]^{l}$" means that each party $P_i$ holds $[\![ x ]\!]_i^{l}$ for $i \in \{0,1\}$.
This scheme ensures information-theoretic privacy of the secret $x$, since each party individually learns nothing about the value of $x$, even with unbounded computational power.

\subsubsection{Oblivious transfer}

In $1$-out-of-$k$ Oblivious Transfer (OT)~\cite{OT}, denoted as $k \choose 1$-OT$_l$, the sender inputs $k$ messages, each with length $l$.
The receiver inputs a value $i$ and learns the message $m_i$, while the sender has no output. 
Utilizing the OT extension technology~\cite{OT-extension}, OT protocols can be efficiently implemented in batches.
This work employs the IKNP-style OT extension~\cite{IKNP}, achieving communication of $2\lambda + kl$ bits in $2$ rounds for ${k \choose 1}$-OT$_l$.
Especially, for ${2 \choose 1}$-OT$_l$, the communication is reduced to $\lambda + 2l$.
The correlated $1$-out-of-$2$ OT~\cite{COT} is a variant of OT, denoted as $2 \choose1$-COT$_l$.
In this functionality, the sender inputs a value $x$, and the receiver inputs a choice bit $b$.
As a result, the sender receives a random value $r$, while the receiver obtains either $r$ or $x + r$, depending on the value of $b$.
The communication cost of ${2 \choose 1}$-COT$_l$ is $\lambda + l$ in $2$ rounds.

\subsubsection{2PC functionalities}

In two-party secret sharing, the expression "$P_0$ and $P_1$ invoke $\mathcal{F}_{f}(x, y)$ to learn $[\![ z ]\!]^l$" means that $P_0$ inputs $x$ and $P_1$ inputs $y$, and they output $[\![ z ]\!]^l$ such that $z = f(x, y)$.
Especially, "Parties input $[\![ a ]\!]$" denotes $P_0$ inputs $[\![ a ]\!]_0$ and $P_1$ inputs $[\![ a ]\!]_1$.
This work utilizes the following functionalities:

\vspace{2pt}
\noindent
\textbf{AND.}
$\mathcal{F}_{\mathsf{AND}}$ allows $P_0$ to input a bit $a$ and $P_1$ to input a bit $b$, such that both parties obtain a Boolean sharing $[\![ c ]\!]^B$ satisfying $c = a \land b$.
This functionality can be implemented using Beaver bit-triples \cite{CrypTFlow2}, with communication $\lambda + 20$ in $2$ rounds.

\vspace{2pt}
\noindent
\textbf{Bit multiplication (BitMul).}
The bit multiplication functionality $\mathcal{F}_{\mathsf{BitMul}}^l$ takes the same inputs as $\mathcal{F}_{\mathsf{AND}}$, but outputs an arithmetic sharing $[\![ c ]\!]^l$ of the result $c = a \land b$.
It can be implemented using one instance of $2 \choose1$-COT$_l$~\cite{geo}, with communication $\lambda + l$ in $2$ rounds.

\vspace{2pt}
\noindent
\textbf{Comparison.}
The comparison functionality $\mathcal{F}_{\mathsf{Comp}}^l$ also known as the Millionaires' functionality, takes $l$-bit $x$ from $P_0$ and $l$-bit $y$ from $P_1$ as input, and outputs a Boolean sharing $[\![ b ]\!]^{B}$ such that $b = \mathsf{Comp}(x,y) = \bm{1}\{x < y\}$.
CrypTFlow2~\cite{CrypTFlow2} implemented an efficient protocol for $\mathcal{F}_{\mathsf{Comp}}^l$, with communication less than $\lambda l + 14l$ in $\log l$ rounds.

\vspace{2pt}
\noindent
\textbf{DReLU.}
The $\mathsf{DReLU}$ functionality $\mathcal{F}_{\mathsf{DReLU}}^l$ takes an arithmetic sharing $[\![ x ]\!]^{l}$ as input, and outputs boolean sharing $[\![ b ]\!]^{B}$ satisfying $b = \mathsf{DReLU}(x) = \bm{1}\{x \in [0, 2^{l-1})\}$.
$\mathcal{F}_{\mathsf{DReLU}}^l$ can be implemented by invoking $\mathcal{F}_{\mathsf{Comp}}^{l-1}$, with communication less than $\lambda (l-1) + 14 (l-1)$ \cite{CrypTFlow2}.

\vspace{2pt}
\noindent
\textbf{Lookup table (LUT).}
The $\mathsf{LUT}$ functionality $\mathcal{F}_{\mathsf{LUT}}^{m,n}$ takes a public or shared table $T$ and an arithmetic sharing index $[\![ I ]\!]^{m}$ as input, where $T$ contains $M =2^m$ entries, each of $n$ bits.
The functionality returns the arithmetic sharing of the $I$-th element in $T$ as output.
$\mathcal{F}_{\mathsf{LUT}}$ can be implemented by invoking a $M \choose 1$-OT$_n$, with communication $2\lambda + Mn$ bits \cite{LUT}.

\vspace{2pt}
\noindent
\textbf{Boolean to Arithmetic (B2A).}
The $\mathcal{F}_{\mathsf{B2A}}^l$ takes $[\![ x ]\!]^B$ as input, and outputs an arithmetic shares $[\![ y ]\!]^l$ such that $x = y$.
It can be implemented by invoking a $2 \choose1$-COT$_{l}$~\cite{CrypTFlow2}, requiring $\lambda + l$ bits of communication in $2$ rounds.

\vspace{2pt}
\noindent
\textbf{Multiplexer (MUX).}
The multiplexer functionality $\mathcal{F}_{\mathsf{MUX}}^l$ takes an arithmetic shares $[\![ x ]\!]^l$ and a Boolean shares $[\![ b ]\!]^B$ as input, and outputs $[\![ y ]\!]^l$ such that $y = x \cdot b$.
$\mathcal{F}_{\mathsf{MUX}}^l$ can be implemented by invoking two $2 \choose 1$-COT$_l$ \cite{CrypTFlow2}, with communication $2(\lambda + l)$.

\vspace{2pt}
\noindent
\textbf{Signed extension (SExt).}
The signed extension functionality $\mathcal{F}_{\mathsf{SExt}}^{l,l'}$ takes $[\![ x ]\!]^l$ as input, and outputs $[\![ y ]\!]^{l'}$ such that $\mathsf{int}(y) = \mathsf{int}(x)$, where $l' > l$.
SirNN~\cite{SiRnn} implements this functionality with communication $\lambda (l + 1) + 13l + l'$.
Further, for the case the input satisfies the constraint $|x| \leq 2^{l-2}$, Guo et al. propose a more efficient implementation \cite{geo}, reduce the communication to $\lambda + l' - l$.

\vspace{2pt}
\noindent
\textbf{Cross term.}
The functionality $\mathcal{F}_{\mathsf{CrossTerm}}^{m,n}$ takes an $m$-bit $x$ from $P_0$ and an $n$-bit $y$ from $P_1$, and outputs $[\![ z ]\!]^{m + n}$ such that $z = xy$.
SirNN \cite{SiRnn} proposes an implementation of this functionality, with communication $u (\lambda + u /2 + 1/2) + mn$, where $u = \min (m,n)$.



\subsubsection{Signed multiplication with non-uniform bitwidths}
\label{subsec:mul}

The signed multiplication with non-uniform bitwidths functionality $\mathcal{F}_{\mathsf{Mul}}^{m,n}$ supports various input formats.
In its general form, $\mathcal{F}_{\mathsf{Mul}}^{m,n}$ takes $[\![ x ]\!]^m$ and $[\![ y ]\!]^n$ as input, and outputs $[\![ z ]\!]^{m + n}$ such that $\mathsf{int}(z) = \mathsf{int}(x) \cdot \mathsf{int}(y)$ \cite{SiRnn}.
This work primarily considers the scenario in which $P_0$ holds an $m$-bit $x$ and $P_1$ holds an $n$-bit $y$.
In this case, the signed product is expressed as: $\mathsf{int}(z) = \mathsf{int}(x) \cdot \mathsf{int}(y) = (x - 2^{m}\cdot \mathsf{MSB}(x)) \cdot (y - 2^{n} \cdot \mathsf{MSB}(y)) = xy - 2^{m} \cdot \mathsf{MSB}(x) \cdot y - 2^{n} \cdot \mathsf{MSB}(y) x + 2^{m + n}\cdot \mathsf{MSB}(x) \cdot \mathsf{MSB}(y)$.
After applying the modulo operation, the last term is zero.
The cross term $xy$ can be computed by invoking a functionality $\mathcal{F}_{\mathsf{CrossTerm}}^{m,n}$.
The remaining two terms can be computed using two instances of the multiplexer functionality $\mathcal{F}_{\mathsf{MUX}}$.
Further, for the special case where $\mathsf{MSB}(x) = \mathsf{MSB}(y) = 0$, we have $\mathsf{int}(z) = x \cdot y$, and only a single invocation of $\mathcal{F}_{\mathsf{CrossTerm}}^{m,n}$ is required.

For the case the $n$-bit $y$ is a public value, both $x$ and $y$ can be (signed) extended to $m+n$ bits using $\mathcal{F}_{\mathsf{SExt}}$, after which the multiplication $x y \bmod 2^{m+n}$ can be computed locally.
Since signed extension is free for public values, the communication of this protocol is equivalent to that of a single $\mathcal{F}_{\mathsf{SExt}}$.




\subsection{Two-party Geometric Method}
\label{sec:pre geo}

\begin{figure}[!htb]
	\centering  
	\subfigbottomskip=2pt 
	\subfigcapskip=-5pt 
	\subfigure [Feasible region for $x \in \mathbb{Z}_{L}$.]{
        \includegraphics[width=0.45\linewidth]{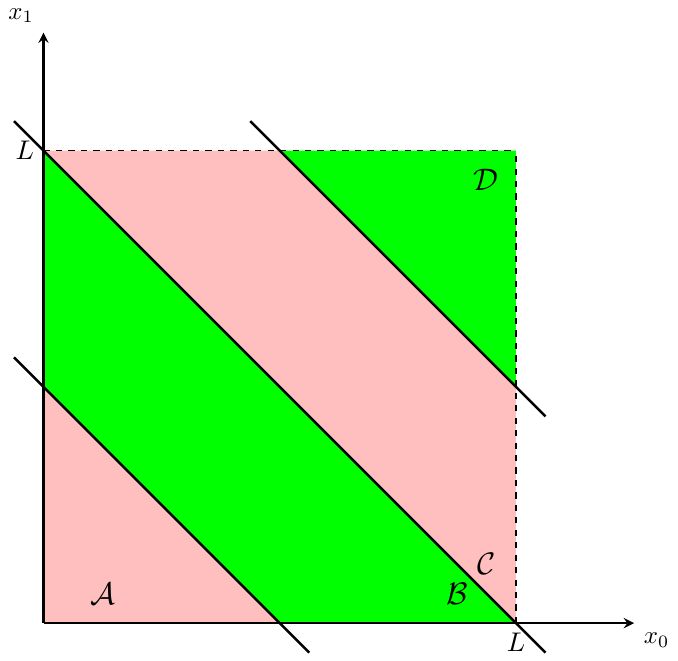}
        \label{fig:MW 0.5N}
        }
	\subfigure[Feasible region for $|x| < \frac{L}{3}$.]{
	   \includegraphics[width=0.45\linewidth]{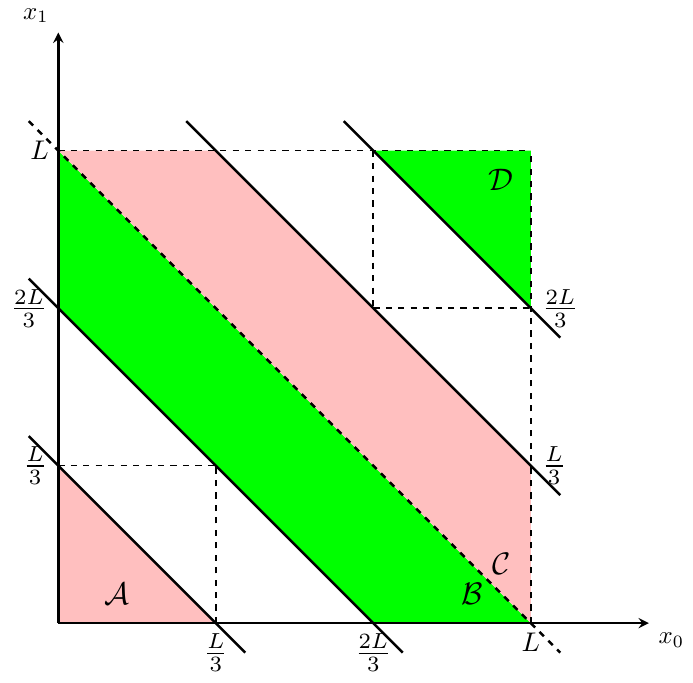}
        \label{fig:MW 0.33N}
        }
        \caption{Feasible region of point $P(x_0, x_1)$.}
        \label{fig:shift}
\end{figure}

For two-party secret sharing $x = x_0 + x_1 \bmod L$, Guo et al. \cite{geo} introduce a MSB-Wrap coefficient $\mathsf{MW}$ defined as
\begin{equation} \label{equ:MW}
    \mathsf{MW}(x_0,x_1, L) = \mathsf{MSB}(x) + \mathsf{Wrap}(x_0, x_1, L),
\end{equation}
where $\mathsf{MSB}(x) = \bm{1}\{x \geq \frac{L}{2}\}$ and $\mathsf{Wrap}(x_0, x_1, L) = \bm{1}\{x_0 + x_1 \geq L\}$.
Moreover, $\mathsf{MW}(x_0,x_1, L)$ is usually abbreviated as $\mathsf{MW}(x, L)$ or $\mathsf{MW}(x)$.
This coefficient $\mathsf{MW}(x)$ enables the computation of the signed integer representation of $x$ as $\mathsf{int}(x) = x_0 + x_1 -\mathsf{MW}(x) \cdot L$.
This representation facilitates direct computation of operations such as truncation, expressed as $x \gg k = \lfloor \frac{\mathsf{int}(x)}{2^k} \rfloor$. 
To compute $\mathsf{MW}(x)$, Guo et al.  propose a novel geometric method.
Specifically, they interpret the shares $x_0$ and $x_1$ as coordinates on a plane, treating secret sharing of $x$ as a point $P(x_0, x_1)$ in the rectangular coordinate system.
For $|x| < B$, where $0 \leq B \leq \frac{L}{2}$, they define four sets as:
\begin{equation}\label{equ:ABCD}
	\begin{split}
		&\mathcal{A}: \{(x_0, x_1) | x_0,x_1 \in \mathbb{Z}_L~,~0 \leq x_0 + x_1 < B \}, \\
		&\mathcal{B}: \{(x_0, x_1) | x_0,x_1 \in \mathbb{Z}_L~,~L - B \leq x_0 + x_1 < L \}, \\
		&\mathcal{C}: \{(x_0, x_1) | x_0,x_1 \in \mathbb{Z}_L~,~L \leq x_0 + x_1 < L + B \}, \\
		&\mathcal{D}: \{(x_0, x_1) | x_0,x_1 \in \mathbb{Z}_L~,~2L - B \leq x_0 + x_1 < 2L \}. \\
	\end{split}
\end{equation}
Then $\mathsf{MSB}(x)$ and $\mathsf{Wrap}(x_0,x_1,L)$ can be computed by determining the area point $P$ falls into, as $\mathsf{MSB}(x) = 1$ if and only if $P \in \mathcal{A} \bigcup \mathcal{C}$, and $\mathsf{Wrap}(x) = 1$ if and only if $P \in \mathcal{C} \bigcup \mathcal{D}$.
Therefore, $\mathsf{MW}(x)$ can be written as: 
\begin{equation}\label{equ:K ABCD}
    \mathsf{MW}(x) = 
    \begin{cases}
        0, \quad \text{if}~ P \in \mathcal{A}, \\
        1, \quad \text{if}~  P \in \mathcal{B} \bigcup \mathcal{C},\\
        2, \quad \text{if}~  P \in \mathcal{D}. \\
    \end{cases}
\end{equation}
as shown in Figure~\ref{fig:MW 0.5N}.

Determining the area in which $P$ falls into is costly for $|x| < \frac{L}{2}$.
However, Guo et al. observe that if $|x| < \frac{L}{3}$, then only two $\mathsf{AND}$ or $\mathsf{BitMul}$ operations are required for determining whether $P \in \mathcal{A}$ or $P \in \mathcal{D}$, as $\bm{1}\{P \in \mathcal{A}\} = \bm{1}\{x_0 < \frac{L}{3}\} \land \bm{1}\{x_1 < \frac{L}{3}\}$ and $\bm{1}\{P \in \mathcal{D}\} = \bm{1}\{x_0 \geq \frac{2L}{3}\} \land \bm{1}\{x_1 \geq \frac{L}{3}\}$ (as shown in Figure~\ref{fig:MW 0.33N}).
Moreover, for the case $|x| < \frac{L}{4}$, only one $\mathsf{BitMul}$ operation is required to compute $\mathsf{MW}(x)$.

\subsection{Threat Model and Security}

This work considers secure two-party computation under the threat model adopted by CrypTFlow2 \cite{CrypTFlow2}.
We assume a static, semi-honest, probabilistic polynomial-time (PPT) adversary that is computationally bounded and may corrupt at most one of the two parties. In the semi-honest model, the corrupted party follows the protocol honestly but attempts to learn private input information from the other party.
We formalize security using the simulation paradigm~\cite{DBLP:journals/joc/Canetti00,DBLP:books/sp/17/Lindell17}. For a given function $f$, we compare two settings: the real interaction and the ideal interaction. 
In the real interaction, the two parties execute the protocol in the presence of an adversary and an external environment $\mathcal{Z}$.
In the ideal interaction, the parties send their inputs to a trusted third party, referred to as the functionality $\mathcal{F}$, which computes $f$ and returns the shares of output to parties.
A protocol is secure if, for every real-world adversary, there exists a simulator $\mathcal{S}$ in the ideal model such that no environment $\mathcal{Z}$ can distinguish between the real and ideal interactions. If such a simulator exists, the real interaction reveals no information about the private inputs, and the protocol is secure.
In practice, a protocol $\Pi$ is often built from multiple sub-protocols, which is described as hybrid model in CrypTFlow2.
The simulation of $\Pi$ follows the structure of a real interaction, except that each sub-protocol is replaced by an invocation of the corresponding ideal functionality. 
When $\Pi$ invokes a functionality $\mathcal{F}$, the protocol is said to operate in the "$\mathcal{F}$-hybrid model".
The security of $\Pi$ then depends on the security of the underlying sub-protocols it calls.

%% file: sec/MW.tex
\section{Compute $\mathsf{MW}(x)$ with Arbitrary Constraint}

According to the definition of $\mathsf{MW}(x)$ in Equation~\ref{equ:MW} or Equation~\ref{equ:MW def1}, computing $\mathsf{MW}(x)$ involves the invocation of comparison protocols.
Moreover, under the constraint $|x| < B$, there exists an additional constraint on the inputs to these comparison protocols. 
In this section, we first give method for performing comparison protocol under special constraint, and subsequently apply this approach to compute $\mathsf{MW}(x)$ for $|x| < B$, with an arbitrary $B \leq \frac{L}{2}$.
Furthermore, for the case that $l$-bit $x$ is shared over $l_r$-bit ring, where $l_r \geq l +1$, we show how to compute $\mathsf{MW}(x \bmod 2^l,2^l)$ from $\mathsf{MW}(x,2^{l_r})$ efficiently.


\subsection{Comparison Protocol with Constraint}
\label{sec:comp}

In general comparison protocol, two participants $P_0$ and $P_1$ hold inputs $x, y \in \mathbb{Z}_{L}$, respectively.
The goal is to compute a comparison function defined as: $\mathsf{Comp}(x,y) = \bm{1}\{x < y\}$.
This work considers a constrained scenario where a specific relationship exists between $x$ and $y$.
Specifically, we assume that $x$ is either less than $y$ or greater than $y$ by a public value $A$, where $A \in (0, L)$.
Formally, the relationship between $x$ and $y$ can be expressed as $x - y \in [A, L) \bigcup [-L, 0)$.
Under this constraint, we present Lemma~\ref{lem:comp A}, which demonstrates that the comparison protocol for inputs $y$ and $x$ can be reduced to the comparison protocol with inputs $\lfloor \frac{y}{A} \rfloor$ and $\lfloor \frac{x}{A} \rfloor$.
Consequently, the input length of the comparison protocol can be reduced from $l$ to $\lceil \log \lfloor \frac{L}{A} \rfloor \rceil$, and the communication complexity is reduced from $O(l)$ to $O(\lceil \log \lfloor \frac{L}{A} \rfloor \rceil)$.
The proof of Lemma~\ref{lem:comp A} is provided in Appendix~\ref{app:proof lem:comp A}.

\begin{lemma}\label{lem:comp A}
    Let $x, y \in \mathbb{Z}_L$ satisfy $x - y \in [A, L) \bigcup [-L, 0)$, where $0 < A < L$.
    Then, $\mathsf{Comp}(y, x) = \mathsf{Comp}(\lfloor \frac{y}{A} \rfloor, \lfloor \frac{x}{A} \rfloor)$.
\end{lemma}

In two-party secret sharing, the comparison protocol is usually used to compute $\mathsf{Wrap}$ function, which can be written as $\mathsf{Wrap}(x_0, x_1, L) = \mathsf{Comp}(L - x_0, x_1)$.
Therefore, for the case the range of $x_0 + x_1$ is constrained as $x_0 + x_1 \in [0, L) \bigcup [L + A, 2L)$, we give Lemma~\ref{lem:wrap A} to compute $\mathsf{Wrap}(x_0, x_1, L)$, whose proof is provided in Appendix~\ref{app:proof lem:wrap A}.

\begin{lemma}\label{lem:wrap A}
    For $x = x_0 + x_1 \bmod L$, if $x_0 + x_1 \in [0, L) \bigcup [L + A, 2L)$, where $0 < A < L$, then $\mathsf{Wrap}(x_0, x_1, L) = \mathsf{Comp}(\lfloor \frac{L - x_0}{A} \rfloor, \lfloor \frac{x_1}{A} \rfloor)$.
\end{lemma}

By applying Lemmas~\ref{lem:comp A} and~\ref{lem:wrap A}, the $\mathsf{Wrap}$ protocol can be realized by a comparison protocol with input length $\lceil \log \lfloor \frac{L}{A} \rfloor \rceil$ instead of $l$, leading to a reduction in communication complexity from $O(l)$ to $O(\lceil \log \lfloor \frac{L}{A} \rfloor \rceil)$.
However, the communication of comparison protocol reported in CrypTFlow2 \cite{CrypTFlow2} is an estimated value, which makes it unsuitable for scenarios with very small input lengths. 
In the following, we present a detailed analysis of the communication complexity for comparison protocols with small input lengths.





\subsubsection{Comparison protocol with very small input length}

The comparison protocol in CrypTFlow2 \cite{CrypTFlow2} relies on $2^m \choose 1$-OT and $\mathsf{AND}$ operations, where $m$ typically set to $4$.
For small input lengths (e.g. $l \leq 4$), the comparison protocol can be implemented using  a single $2^l \choose 1$-OT, with communication $2\lambda + 2^l$.
Additionally, CrypTFlow2's comparison protocol outputs a Boolean sharing $[\![ b ]\!]^{B}$, and an extra $\mathsf{B2A}$ protocol is required to convert it to arithmetic sharing $[\![ b ]\!]^{l'}$.
Therefore, the total communication of the comparison protocol with arithmetic sharing output is $3\lambda + 2^l + l'$ when $l \leq 4$.

For very small inputs, an $\mathsf{AND}$-based method can also be used to implement the comparison protocol.
Moreover, we can use $\mathsf{BitMul}$ instead of $\mathsf{AND}$ to directly output arithmetic sharing.
Specifically, suppose $P_0$ and $P_1$ hold inputs $x$ and $y$, where $x, y \in \mathbb{Z}_{n}$.
The comparison function can be computed as:
$
    \mathsf{Comp}(x, y) = \sum_{i=0}^{n-2} \bm{1}\{x = i\} \land \bm{1}\{y > i\}.
$
This requires $n - 1$ $\mathsf{BitMul}$ protocols, directly producing $[\![ b ]\!]^{l'}$, with a total communication cost $(n-1) \cdot (\lambda + l')$ in $2$ rounds.
When comparing these two methods, we find that the $\mathsf{AND}$-based method is more efficient when $n \leq 3$.
Notably, the method in \cite{geo} can be regarded as an instance of the $\mathsf{AND}$-based comparison protocol.

\subsection{Computing $\mathsf{MW}(x)$ with Constraint}
\label{sec: subsection MW}

\begin{figure}[!htb]
	\centering  
	\subfigbottomskip=2pt 
	\subfigcapskip=-5pt 
	\subfigure [Feasible region of $P(x_0,x_1)$.]{
        \includegraphics[width=0.48\linewidth]{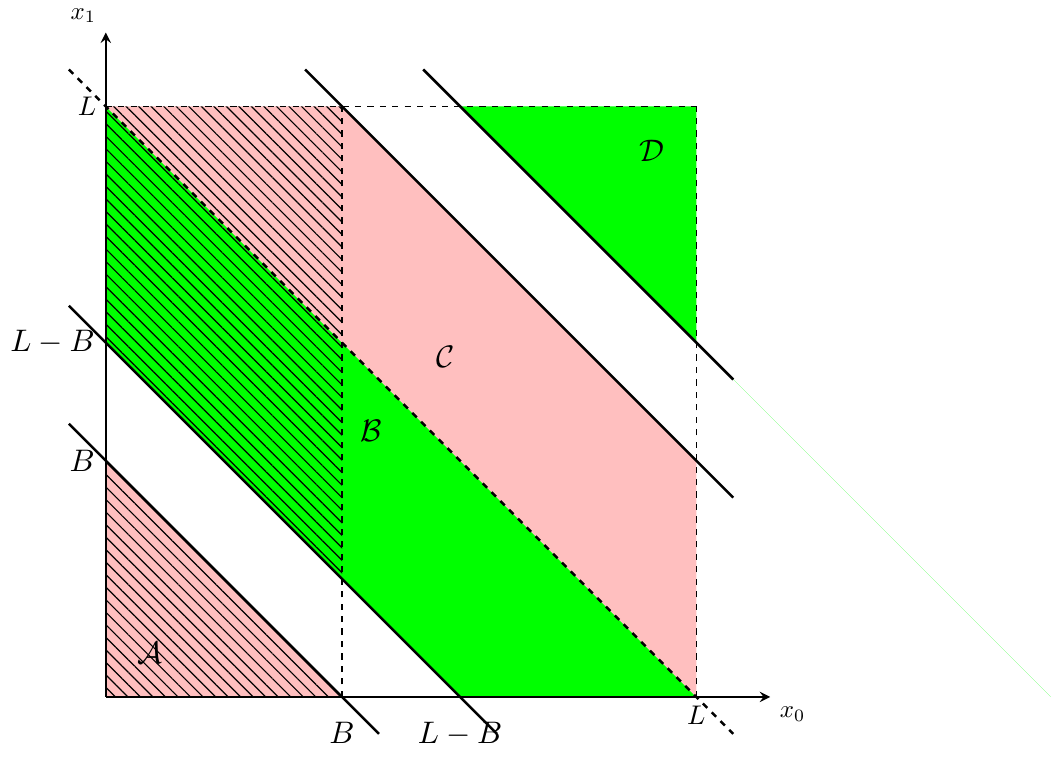}
        \label{fig:2-p feasible region}
        }
	\subfigure[Feasible region of $P^*(x_0^*,x_1)$.]{
	   \includegraphics[width=0.42\linewidth]{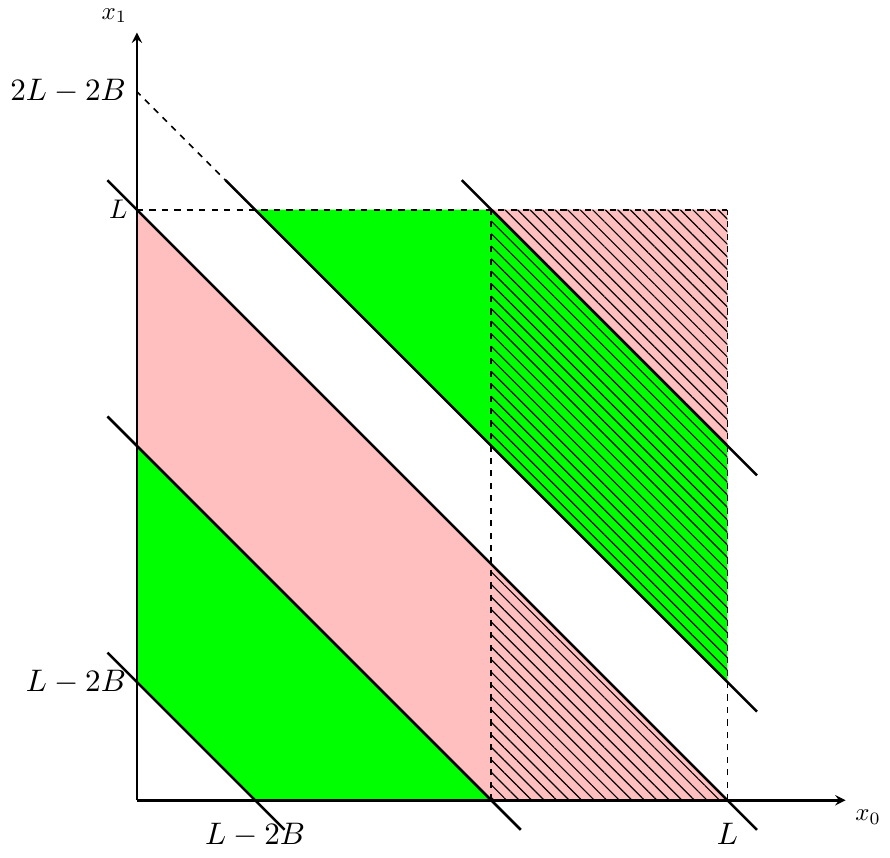}
        \label{fig:2-party-shift-origin}
        }
        \caption{Feasible region of $P(x_0,x_1)$ and $P^*(x_0^*,x_1)$.}
        \label{fig:shift}
\end{figure}

From the definition of $\mathsf{MW}(x)$ in Equation~\ref{equ:MW}, two comparison protocols are required to compute $\mathsf{MSB}(x)$ and $\mathsf{Wrap}(x_0,x_1, L)$, respectively.
However, using the idea of Theorem 2 in \cite{geo}, we can reduce the number of comparison protocols to one.
Specifically, consider the case where $|x| < B$ for $B \leq \frac{L}{2}$.
The feasible region of the point $P(x_0, x_1)$ is illustrated in Figure~\ref{fig:2-p feasible region}. By shifting this feasible region to the left by $B$, we define a new point $P^{*}(x_0^*, x_1)$ where $x_0^* = x_0 - B \bmod L$.
The feasible region for $P^*$ is depicted in Figure~\ref{fig:2-party-shift-origin}.
We introduce a new variable $M^* = \bm{1}\{x_0^* + x_1 \geq 2L - 2B\}$, which indicates whether $P^*$ lies within the upper-right triangular region in Figure~\ref{fig:2-party-shift-origin}. 
Using this definition, $\mathsf{MW}(x)$ can be computed based on $M^*$, as formalized in Theorem~\ref{theo:MW}. The proof of this theorem is provided in Appendix~\ref{app:proof theo:MW}.

\begin{theorem}\label{theo:MW}
Let $x = x_0 + x_1 \bmod L$ and $|x| < B$, where $B \leq \frac{L}{2}$.
Define $x_0^* = x_0 - B \bmod L$, $M^* = \bm{1}\{x_0^* + x_1 \geq 2L - 2B\}$ and $\Delta = \bm{1}\{x_0 \geq B\}$.
Then $\mathsf{MW}(x) = M^* + \Delta$.
\end{theorem}

Notably, Theorem 2 in \cite{geo} is a special case of Theorem~\ref{theo:MW} in this work, where they set $B = \frac{L}{4}$.
Furthermore, Theorem~\ref{theo:MW} also holds for $B = \frac{L}{2}$, enabling its application to any $x \in \mathbb{Z}_L$ without constraint.
In Theorem~\ref{theo:MW}, $\Delta$ can be computed locally by $P_0$.
Thus, the primary challenge in computing $\mathsf{MW}(x)$ lies in determining $M^* = \bm{1}\{x_0^* + x_1 \geq 2L - 2B\}$.
For $|x| < B$, the value of $x_0^* + x_1$ is constrained to the range $[L-2B,L) \bigcup [2L-2B, 2L)$, as shown in Figure~\ref{fig:2-party-shift-origin}. 
If $B = \frac{L}{2}$, then $\mathsf{Wrap}(x_0^*, x_1, L) = \mathsf{Comp}(x_0^*, L - x_1) \oplus 1$.
For the case $B \neq \frac{L}{2}$, from Lemma~\ref{lem:wrap A}, we can derive that 
$M^* = \mathsf{Wrap}(x_0^*, x_1, L) = \mathsf{Comp}(\lfloor \frac{L - x_0^*}{L-2B} \rfloor, \lfloor \frac{x_1}{L - 2B} \rfloor)$.
Under the constraint $|x| < B$, defining $l^* = \lceil \log \lfloor \frac{L}{L - 2B} \rfloor \rceil$ for $B \neq \frac{L}{2}$, and $l^* = l$ when $B = \frac{L}{2}$, comparison protocol with $l^*$-bit input is required for computing $M^*$ and $\mathsf{MW}(x)$.
The detailed procedure for computing $\mathsf{MW}(x)$ is outlined in Algorithm~\ref{alg:MW}.

In this Algorithm~\ref{alg:MW}, the comparison protocol can be implemented using either the $\mathsf{AND}$-based method or CrypTFlow2's method, depending on which achieves the best efficiency. 
Specifically, let the maximum input value of comparison protocol for computing $M^*$ be defined as $K = \lfloor \frac{L}{L-2B} \rfloor$.
The $\mathsf{AND}$-based method is more efficient for computing $M^*$ when $K \leq 3$, which implies $B < \frac{3L}{8}$.
Under this constraint, computing $M^*$ requires only $K$ $\mathsf{BitMul}$ protocols.
For the case $B \geq \frac{3L}{8}$, the CrypTFlow2's comparison protocol is invoked.

\begin{algorithm}[!htb]
	\caption{Computing $\mathsf{MW}(x)$ for $|x| < B$, $\Pi_{\mathsf{MW}}^{l, l'}$: }
	\label{alg:MW}
	\LinesNumbered 

	\KwIn{$P_0$ and $P_1$ hold $[\![ x ]\!]^{l}$ satisfying $|x| < B$, where $B \leq \frac{L}{2}$.}
 
	\KwOut{$P_0$ and $P_1$ output $[\![ \mathsf{MW}(x) ]\!]^{l'}$.}

    Let $L = 2^l$, $K = \lfloor \frac{L}{L-2B} \rfloor$, $l^* = \lceil \log \lfloor \frac{L}{L - 2B} \rfloor \rceil$ if $B < \frac{L}{2}$, and $l^* = l$ if $B = \frac{L}{2}$.
    
    $P_0$ computes $\delta = \bm{1}\{x_0 \geq B\}$ and $x_0^* = x_0 - B \bmod L$.

    \tcp {CrypTFlow2-based Comparison Protocol}
    \If{$B \geq \frac{3L}{8}$}
    {
    
        \If{$B = \frac{L}{2}$}
        {
        $P_0$ and $P_1$ invoke $\mathcal{F}_{\mathsf{Comp}}^{l^*}( x_0^*, L - x_1)$ and learn $[\![temp ]\!]^{B}$.

        $P_0$ and $P_1$ compute $[\![ M^{*} ]\!]^{B} = [\![ temp ]\!]^{B} \oplus 1$.

        }
        \Else
        {
        $P_0$ and $P_1$ invoke $\mathcal{F}_{\mathsf{Comp}}^{l^*}(\lfloor \frac{L - x_0^*}{L-2B} \rfloor, \lfloor \frac{x_1}{L-2B} \rfloor)$ and learn $[\![ M^{*} ]\!]^{B}$.
        }

        $P_0$ and $P_1$ invoke $\mathcal{F}^{l'}_{\mathsf{B2A}}([\![ M^{*} ]\!]^{B})$ and learn $[\![ M^{*} ]\!]^{l'}$.
    }
    
    \tcp {AND-based Comparison Protocol}
    \If{$B < \frac{3L}{8}$}
    {
        \For{$i \in \{0,..., K-1\}$} 
        {
            $P_0$ sets $a_i = \bm{1}\{\lfloor \frac{L - x_0^*}{L - 2B} \rfloor = i\}$ and $P_1$ sets $b_i = \bm{1}\{\lfloor \frac{x_1}{L - 2B} \rfloor > i \}$.

            $P_0$ and $P_1$ invoke $\mathcal{F}_{\mathsf{BitMul}}^{l'}(a_i, b_i)$ to learn $[\![ c_i ]\!]^{l'}$.

        }

        $P_0$ and $P_1$ compute $[\![ M^* ]\!]^{l'}= \sum_{i=0}^{K-1}[\![ c_i ]\!]^{l'}$.
        
    }

    $P_0$ and $P_1$ compute and output $[\![ MW ]\!]^{l'} = [\![ M^* ]\!]^{l'} + \delta$.

\end{algorithm}

\vspace{5pt}
\noindent
\textbf{Correctness and security.}
The correctness of $\Pi_{\mathsf{\mathsf{MW}}}$ is guaranteed by Theorem~\ref{theo:MW}.
The security of $\Pi_{\mathsf{MW}}$ is ensured by the security of the underlying protocols for functionalities 
$\mathcal{F}_{\mathsf{BitMul}}$, 
$\mathcal{F}_{\mathsf{Comp}}$, and
$\mathcal{F}_{\mathsf{B2A}}$.

\vspace{5pt}
\noindent
\textbf{Complexity.}
For $B < \frac{3L}{8}$, $\Pi_{\mathsf{\mathsf{MW}}}$ invokes $K$ instances of the $\mathcal{F}_{\mathsf{BitMul}}^{l'}$, resulting in a total communication cost of $K \cdot (\lambda + l')$ in two rounds.
For $B \geq \frac{3L}{8}$, the protocol requires one $\mathcal{F}_{\mathsf{Comp}}^{l^*}$ and one $\mathcal{F}_{\mathsf{B2A}}^{l'}$, with a total communication cost of $(l^*+1) \lambda + 14l^* + l'$ in $2 + \log l^*$ rounds.

\subsection{Computing $\mathsf{MW}$ via Ring Conversion}
\label{sec: subsection MW ring conversion}

Suppose a protocol $\Pi_{f}$ consisting of two parts: computing $\mathsf{MW}(x)$ and other operations, with their respective communication complexity denoted as $C_{\mathsf{MW}}$ and $C_{\mathsf{other}}$, where $C_{\mathsf{other}}$ scales linearly with the input length.
Consider the scenario that an $l$-bit value $x$ is shared over ring $\mathbb{Z}_{2^{l_r}}$, where $l_r \geq l + 1$.
To evaluate $f(x)$, the straightforward method is to perform $\Pi_{f}$ over the ring $\mathbb{Z}_{2^{l_r}}$, resulting $C_{\mathsf{MW}} \approx \lambda$ as $|x| < 2^{l_r - 2}$, while $C_{\mathsf{other}}$ is $O({l_r})$.
One can also first compute $z_i = x_i \bmod 2^l$ for $i \in \{0,1\}$ and $z = z_0 + z_1 \bmod 2^l$, and perform $\Pi_{f}$ with $[\![ z ]\!]^{l}$ as input.
This reduces $C_{\mathsf{other}}$ to $O(l)$, while $C_{\mathsf{MW}}$ increases to $O(l)$, as the $l$-bit comparison protocol is required to compute $\mathsf{MW}(x, 2^l)$.
To benefit from the advantages of both methods, we propose a technique to compute $\mathsf{MW}(z, 2^l)$ from $\mathsf{MW}(x, 2^{l_r})$ with low overhead, reducing the total communication complexity of $\Pi_{f}$ to $\lambda + O({l})$.






\subsubsection{Computing $\mathsf{MW}(z, 2^{l})$ from $\mathsf{MW}(y, 2^{l + 1})$}

\begin{figure}[!htb]
	\centering  
	\subfigbottomskip=2pt 
	\subfigcapskip=-5pt 
	\subfigure [Feasible region of $P_y(y_0. y_1)$.]{
        \includegraphics[width=0.45\linewidth]{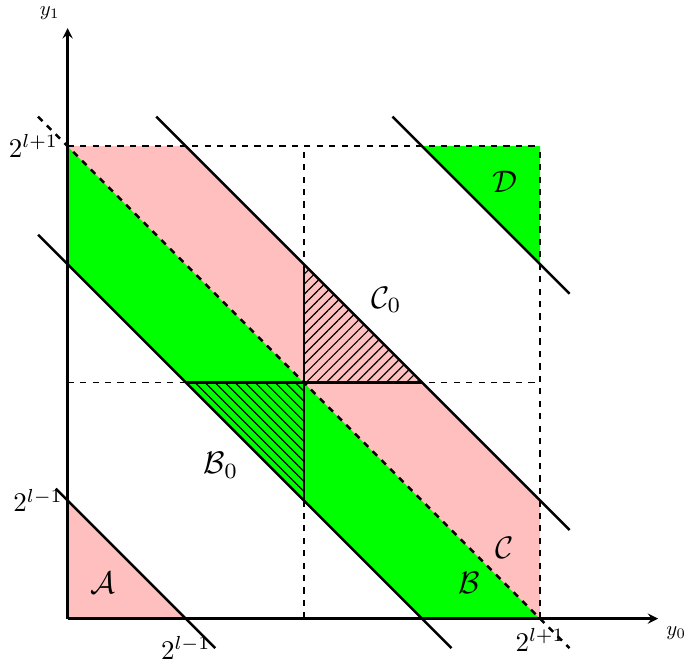}
        \label{fig:MW_conv_before}
        }
	\subfigure[Feasible region of $P_z(z_0, z_1)$.]{
	   \includegraphics[width=0.45\linewidth]{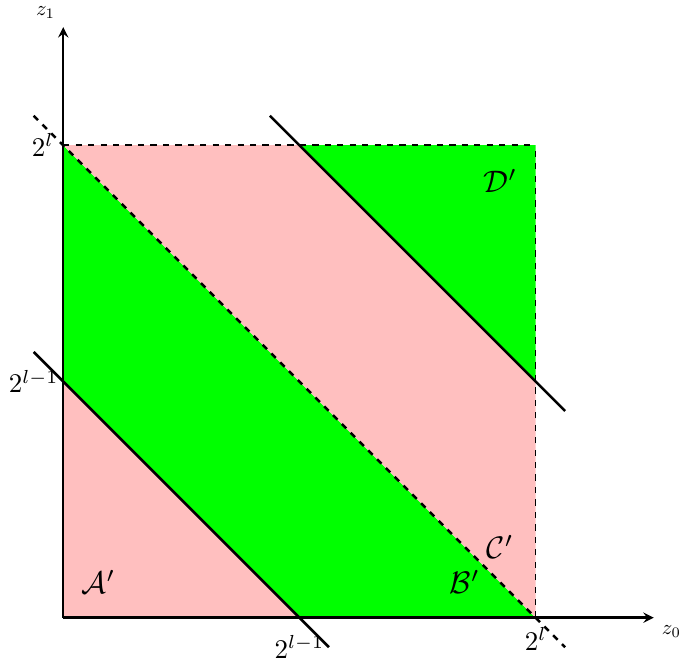}
        \label{fig:MW_conv_after}
        }
        \caption{Feasible region of $P_y(y_0, y_1)$ and $P_z(z_0, z_1)$, where $y \in\mathbb{Z}_{2^l+1}$ and $z \in \mathbb{Z}_{2^{l}}$.
        }
        \label{fig:MW_conv}
\end{figure}

For $x = x_0 + x_1 \bmod 2^{l_r}$, where $|x| < 2^{l-1}$, let $y_i = x_i \bmod 2^{l + 1}$ for $i \in \{0,1\}$ and $y = y_0 + y_1 \bmod 2^{l + 1}$.
Then $\mathsf{int}(x) = \mathsf{int}(y)$, $|y| < 2^{l-1}$, and $\mathsf{MW}(y, 2^{l +1})$ can be easily computed as $|y| < \frac{2^{l+1}}{4}$.
Let $z_i = x_i \bmod 2^l$ for $i \in \{0,1\}$ and $z = z_0 + z_1 \bmod 2^l$, we show how to compute $\mathsf{MW}(z, 2^l)$ from $\mathsf{MW}(y, 2^{l +1})$.
Using the geometric method in \cite{geo}, in addition to the feasible regions $\mathcal{A}$, $\mathcal{B}$, $\mathcal{C}$ and $\mathcal{D}$ defined in Equation~\ref{equ:ABCD}, we introduce two additional regions $\mathcal{B}_0$ and $\mathcal{C}_0$, for $x \in \mathbb{Z}_N$, defined as:
\begin{equation*}\label{equ:B0C0}
	\begin{split}
        &\mathcal{B}_0: \{(x_0, x_1) | (x_0,x_1) \in \mathcal{B}~,~x_0 < \frac{N}{2}, x_1 < \frac{N}{2} \}, \\
        &\mathcal{C}_0: \{(x_0, x_1) | (x_0,x_1) \in \mathcal{C}~,~x_0 \geq \frac{N}{2}, x_1 \geq \frac{N}{2} \}.
	\end{split}
\end{equation*}
Then, the feasible region of point $P_y(y_0, y_1)$ ($y \in\mathbb{Z}_{2L}$) is illustrated in Figure~\ref{fig:MW_conv_before}, where in this figure the value of $N$ is $2L$.
By mapping point $P(x_0, x_1)$ to point $P_z(z_0, z_1)$, where $z_i = x_i \bmod 2^l$ for $i \in \{0,1\}$, the feasible region of point $P_z$ is shown in Figure~\ref{fig:MW_conv_after}.
This region is divided into four sub-regions, denoted as $\mathcal{A}'$, $\mathcal{B}'$, $\mathcal{C}'$ and $\mathcal{D}'$.
From these two figures, the relationship between points $P_y(y_0, y_1)$ and $P_z(z_0, z_1)$ can be summarized as follows:
\begin{itemize}
    \item 
    If $P_y \in \mathcal{A} \bigcup \mathcal{C}_0$, then $P_z \in \mathcal{A}'$.
    
    \item 
    If $P_y \in \mathcal{D} \bigcup \mathcal{B}_0$, then $P_z \in \mathcal{D}'$.

    \item 
    If $P_y \in (\mathcal{B} \bigcup \mathcal{C}) \setminus (\mathcal{B}_0 \bigcup \mathcal{C}_0)$, then $P_z \in \mathcal{B}' \bigcup \mathcal{C}'$.
\end{itemize}
Thus, we can conclude that $\mathsf{MW}(y, 2^{l+1}) = \mathsf{MW}(z, 2^l)$, except in cases where $P_y \in \mathcal{C}_0 \bigcup \mathcal{B}_0$.
Furthermore, whether this exception occurs can be determined using two $\mathsf{AND}$ operations.
This allows us to compute $\mathsf{MW}(z, 2^l)$ from $\mathsf{MW}(y, 2^{l+1})$.
The following Lemma~\ref{lem:MW conv 1} formalizes this proposition, with its proof provided in Appendix~\ref{app:proof lem:MW conv 1}.

\begin{lemma}\label{lem:MW conv 1}
    Suppose $y = y_0 + y_1 \bmod 2^{l+1}$ satisfying $|y| < 2^{l-1}$.
    For $i \in \{0,1\}$, let $z_i = y_i \bmod 2^l$ and $z = z_0 + z_1 \bmod 2^l$.
    Define:
    \begin{equation}\label{equ: a b}
        \begin{cases}
            a = \bm{1}\{y_0 \in [2^{l}, 2^{l} + 2^{l-1})\} \land \bm{1}\{y_1 \in [2^{l}, 2^{l} + 2^{l-1})\}, \\
            b = \bm{1}\{y_0 \in [2^{l} - 2^{l-1} , 2^{l})\} \land \bm{1}\{y_1 \in [2^{l} - 2^{l-1}, 2^{l})\}.
        \end{cases}
    \end{equation}
    Then $\mathsf{MW}(z, 2^l) = \mathsf{MW}(y, 2^{l+1}) - a + b$.
\end{lemma}

\subsubsection{Reducing the number of AND operations to one}

For given $\mathsf{MW}(y, 2^{l+1})$, only $a - b$ is required to compute $\mathsf{MW}(z,2^{l})$, which typically involves two additional $\mathsf{AND}$ operations.
Now, we demonstrate how to compute $a - b$ using only a single $\mathsf{AND}$ operation.
From Equation~\ref{equ: a b} we have that $a = \bm{1}\{P_y \in \mathcal{C}_0\}$, and $b = \bm{1}\{P_y \in \mathcal{B}_0\}$.
For point $P_y$ in Figure~\ref{fig:MW_conv_before}, we map $P_y(y_0, y_1)$ to $\hat{P}_y(\hat{y}_0, \hat{y}_1)$, where $\hat{y}_i = y_i + 2^{l} \bmod 2^{l+1}$, and let $\hat{y} = \hat{y}_0 + \hat{y}_1 \bmod 2^{l+1}$.
Then in Figure~\ref{fig:MW_conv_before}, we can deduce that:
\begin{itemize}
    \item 
    $P_y \in \mathcal{C}_0$ (or equivalently, $a = 1$) if and only if $\hat{P}_y \in \mathcal{A}$ (or equivalently, $\mathsf{MW}(\hat{y}, 2^{l+1}) = 0$).

    \item 
    $P_y \in \mathcal{B}_0$ (or equivalently, $b = 1$) if and only if $\hat{P}_y \in \mathcal{D}$ (or equivalently, $\mathsf{MW}(\hat{y}, 2^{l+1}) = 2$).
\end{itemize}
Therefore, computing $a - b$ reduces to evaluating $\mathsf{MW}(\hat{y}, 2^{l+1})$, which requires only a single $\mathsf{AND}$ operation as $|y| < 2^{l-1}$.
The formal explanation is provided in Theorem~\ref{lem:MW conv 2}, with its proof detailed in Appendix~\ref{app:proof lem:MW conv 2}.

\begin{theorem}\label{lem:MW conv 2}
    Suppose $y = y_0 + y_1 \bmod 2^{l+1}$, where  $|y| < 2^{l-1}$, 
    and $a$, $b$ are defined as Equation~\ref{equ: a b}.
    Let $\hat{y}_i = y_i + 2^{l} \bmod 2^{l+1}$ for $i \in \{0,1\}$, $\hat{y}^*_0 = \hat{y}_0 - 2^{l-1} \bmod 2^{l+1}$, and $\delta = \bm{1}\{\hat{y}_0 \geq 2^{l - 1}\}$.
    Define $\hat{M}^* = \bm{1}\{\hat{y}_0^* \geq 2^{l}\} \land \bm{1}\{\hat{y}_1 \geq 2^{l}\}$.
    Then $a - b = 1 - \delta - \hat{M}^*$.
\end{theorem}

Using Lemma~\ref{lem:MW conv 1} and Theorem~\ref{lem:MW conv 2}, $\mathsf{MW}(z, 2^{l})$ can be computed from $\mathsf{MW}(y,2^{l+1})$ with only one additional $\mathsf{AND}$ operation.
The details for computing $\mathsf{MW}(z, 2^{l})$ from $x \in \mathbb{Z}_{2^{l_r}}$ is shown in Algorithm\ref{alg:MW conv}.

\begin{algorithm}[!htb]
	\caption{Computing $\mathsf{MW}(z, 2^{l})$ from $x \in \mathbb{Z}_{2^{l_r}}$,  $\Pi_{\mathsf{MW_{conv}}}^{l_r, l'}$: }
	\label{alg:MW conv}
	\LinesNumbered 

	\KwIn{$P_0$ and $P_1$ hold $[\![ x ]\!]^{l_r}$, where $|x| < 2^{l-1}$ and $l_r \geq l + 1$.}

	\KwOut{$P_0$ and $P_1$ output $[\![ \mathsf{MW}_z ]\!]^{l'}$, where $\mathsf{MW}_z = \mathsf{MW}(z,2^l)$. $z$ is shared over $\mathbb{Z}_{2^l}$ and $[\![ z ]\!]^{l}_i = [\![ x ]\!]^{l_r}_i \bmod 2^l$ for $i \in \{0,1\}$.}

    For $i \in \{0,1\}$, $P_i$ computes $[\![ y ]\!]^{l+1}_i = x_i \bmod 2^{l+1}$.

    \tcp {Compute MW(y)}

    $P_0$ and $P_1$ invoke $\mathcal{F}_\mathsf{MW}^{l+1, l'}([\![ y ]\!]^{l+1})$ to learn $[\![ \mathsf{MW}_y ]\!]^{l'}$.

    \tcp {Compute a - b}

    For $i \in \{0,1\}$, $P_i$ computes $\hat{y}_i = [\![ y ]\!]^{l+1}_i + 2^l \bmod 2^{l+1}$.

    $P_0$ computes $\hat{y}_0^* = \hat{y}_0 - 2^{l-1} \bmod 2^{l+1}$, and 
    \textcolor{black}{$\delta = \bm{1}\{\hat{y}_0 \geq 2^{l - 1}\}$}
    .

    $P_0$ and $P_1$ invoke $\mathcal{F}_{\mathsf{BitMul}}^{l'}(\bm{1}\{\hat{y}_0^* \geq 2^{l}\}, \bm{1}\{\hat{y}_1 \geq 2^{l}\})$ and learn $[\![ \hat{M}^* ]\!]^{l'}$.

    $P_0$ and $P_1$ compute 
    \textcolor{black}{$[\![ c ]\!]^{l'} = 1 - \delta - [\![ \hat{M}^* ]\!]^{l'}$}
    .

    \tcp {Compute MW(z)}
    
    $P_0$ and $P_1$ compute $[\![ \mathsf{MW}_z ]\!]^{l'} = [\![ \mathsf{MW}_y ]\!]^{l'}  - [\![ c ]\!]^{l'}$.
    
    $P_0$ and $P_1$ output $[\![ \mathsf{MW}_z ]\!]^{l'}$.

\end{algorithm}

\vspace{5pt}
\noindent
\textbf{Correctness and security.}
The correctness of $\Pi_{\mathsf{MW_{conv}}}^{l_r, l'}$ is ensured by Lemma~\ref{lem:MW conv 1} and Theorem~\ref{lem:MW conv 2}.
The security comes from the security of protocols for 
$\mathcal{F}_{\mathsf{MW}}$ and 
$\mathcal{F}_{\mathsf{BitMul}}$.

\vspace{5pt}
\noindent
\textbf{Complexity.}
The communication of $\Pi_{\mathsf{MW_{conv}}}^{l_r, l'}$ arises from the invocation of $\mathcal{F}_{\mathsf{MW}}^{l+1, l'}$ and $\mathcal{F}_{\mathsf{BitMul}}^{l'}$.
In $\mathcal{F}_{\mathsf{MW}}^{l+1, l'}$, the constraint is $|y| < \frac{2^{l+1}}{4}$, resulting $K = 1$ in Algorithm~\ref{alg:MW}, and only one $\Pi_{\mathsf{BitMul}}$ is required, with communication $\lambda + l'$.
Therefore, the total communication of $\Pi_{\mathsf{MW_{conv}}}^{l_r, l'}$ is $2(\lambda + l')$.

%% file: sec/APP.tex
\section{Application on Real-World Functions}
\label{sec:app all}


This work focuses on evaluating a function $\mathsf{func}(\cdot)$ with the form of Equation~\ref{equ:func x y z}.
Once $\mathsf{MW}(x)$ is obtained, Equation~\ref{equ:func int} can be used to compute $\mathsf{func}(\cdot)$, following the approach outlined in Section~\ref{subsec:app}.
The evaluation process proceeds as follows:
\begin{enumerate}
    \item 
    For $i \in \{0,...,k-1\}$, $P_0$ and $P_1$ locally compute $f_i(\frac{x_0}{2^f})$, and $g_i(\frac{x_1}{2^f})$, respectively.
    Then they compute public functions $h_i(\frac{- j \cdot L}{2^f})$ for $j \in \{0,1,2\}$.

    \item 
    For $i \in \{0,...,k-1\}$, parties invoke multiplication protocol to compute $z_i = f_i(\frac{x_0}{2^f}) \cdot g_i(\frac{x_1}{2^f})$.

    \item 
    For $i \in \{0,...,k-1\}$, and $j \in \{0,1,2\}$ parties compute $t^j_i = z_i \cdot h_i(\frac{- j \cdot L}{2^f})$, and store them into a lookup table $T_i$.

    \item 
    $P_0$ and $P_1$ invoke $\mathcal{F}_{\mathsf{MW}}$ to get $\mathsf{MW}(x)$.

    \item 
    For $i \in \{0,...,k-1\}$, $P_0$ and $P_1$ query the lookup table $T_i$ with index $\mathsf{MW}(x)$ to get $t^{\mathsf{MW}}_i$.

    \item 
    $P_0$ and $P_1$ locally compute and output $\sum_i^k t_i$.
    
\end{enumerate}
This method requires only a limited number of multiplications, a computation of $\mathsf{MW}$, and a lookup table protocol.
In the following subsections, we demonstrate the applicability of this approach to real-world functions.

\subsection{Division and Truncation Functions}
\label{sec:app div}

The division function can be expressed in the form of Equation~\ref{equ:func x y z} as $\frac{a+b+c}{d} = \frac{a}{d} \cdot 1 \cdot 1 + 1 \cdot \frac{b}{d} \cdot 1 + 1 \cdot 1 \cdot \frac{c}{d}$.
This work adopts the definition of signed integer division in CrypTFlow2 \cite{CrypTFlow2}.
Given a shared input $x$, and a public divisor $d$ encoded on ring $\mathbb{Z}_{L}$, the integer division function can be written as: 
\begin{equation}\label{equ:division}
    \begin{split}
        \mathsf{Div}(x,d) &= \lfloor \frac{\mathsf{int}(x)}{d} \rfloor
        = \lfloor \frac{x_0 + x_1 - \mathsf{MW}(x) \cdot L}{d} \rfloor \\
        &= \lfloor \frac{x_0}{d} \rfloor + \lfloor \frac{x_1}{d} \rfloor - \mathsf{MW}(x) \cdot \lfloor \frac{L}{d} \rfloor + e, 
    \end{split}
\end{equation}
where $e$ is a small error resulted from the floor operations.
To realize faithful division protocol, the error $e$ should be computed and eliminated.
We reformulate Equation~\ref{equ:division} as: 
\begin{equation}\label{equ:division faithful}
    \mathsf{Div}(x,d)=\lfloor \frac{\mathsf{int}(x)}{d} \rfloor 
    = \lfloor \frac{x_0}{d} \rfloor + \lfloor \frac{x_1 - \mathsf{MW}(x) \cdot L}{d}  \rfloor + \epsilon,
\end{equation}
where $\epsilon \in \{0,1\}$ is an one-bit error.
According to Lemma 2 from \cite{geo}, $\epsilon$ can be computed as:
\begin{equation}\label{equ:epsilon}
\begin{split}
    \epsilon = \bm{1}\{x_0 \bmod d + (x_1 - \mathsf{MW}(x) \cdot L \bmod d) \geq d\}.
\end{split}
\end{equation}
The remaining challenge lies in computing the terms $\lfloor \frac{x_1 - \mathsf{MW}(x) \cdot L}{d} \rfloor$ and $(x_1 - \mathsf{MW}(x) \cdot L) \mod d$, which can be computed using lookup table (LUT) technology.
Specifically, we construct two lookup tables, $T^{d}$ and $T^{\epsilon}$, where $T^{d}[j] = \lfloor \frac{x_1 - j \cdot L}{d} \rfloor$ and $T^{\epsilon}[j] = (x_1 - j \cdot L) \mod d$ for $j \in \{0,1,2\}$.
These tables are indexed by $\mathsf{MW}(x)$, allowing  parties to retrieve the required values efficiently.
Finally, the parties invoke the $\mathsf{DReLU}$ protocol to compute the error $\epsilon$.
The details of the division protocol are shown in Algorithm~\ref{alg:division faithful}.

\begin{algorithm}[!htb]
	\caption{Faithful division protocol, $\Pi_{\mathsf{Div}}^{l,d}$: }
	\label{alg:division faithful}
	\LinesNumbered 

	\KwIn{$P_0$ and $P_1$ hold $[\![ x ]\!]^{l}$ and a public $d$, satisfying $|x| < B$ where $B \leq \frac{L}{2}$.}
 
	\KwOut{$P_0$ and $P_1$ output $[\![ \lfloor \frac{\mathsf{int}(x)}{d} \rfloor ]\!]^{l}$.}

    Let $l_d = \lceil \log d \rceil$.
    
    \For{$j \in \{0,1,2\}$}
    {

        $P_0$ sets $[\![T^{d} ]\!]_0^{l}[j] = 0$ and $ [\![T^{\epsilon} ]\!]_0^{l_d + 1}[j] = 0$.
    
        $P_1$ sets $[\![T^{d}]\!]_1^{l}[j] = \lfloor \frac{x_1 - j \cdot L}{d} \rfloor$ and $[\![T^{\epsilon}]\!]_1^{l_d+1}[j] = (x_1 - j \cdot L) \mod d$.
    }

    $P_0$ and $P_1$ invoke $\mathcal{F}_{\mathsf{MW}}^{l,2}([\![ x ]\!]^{l})$ and learn $[\![ \mathsf{MW} ]\!]^{2}$.

    $P_0$ and $P_1$ invoke $\mathcal{F}_{\mathsf{LUT}}([\![T^{d}]\!]^l, [\![ \mathsf{MW} ]\!]^{2})$ to learn $[\![ X_1 ]\!]^{l}$.

    $P_0$ and $P_1$ invoke $\mathcal{F}_{\mathsf{LUT}}([\![T^{\epsilon}]\!]^{l_d+1}, [\![ \mathsf{MW} ]\!]^{2})$ to learn $[\![ I_\epsilon ]\!]^{l_d+1}$.

    $P_0$ and $P_1$ compute $[\![ temp ]\!]^{l_d+1} = (x_0 \mod d) + [\![ I_\epsilon ]\!]^{l_d+1} - d$.

    $P_0$ and $P_1$ invoke $\mathcal{F}_{\mathsf{DReLU}}^{l_d + 1}([\![ temp ]\!]^{l_d+1})$ to learn $[\![ \epsilon ]\!]^{B}$.

    $P_0$ and $P_1$ invoke $\mathcal{F}_{\mathsf{B2A}}^{l}([\![ \epsilon ]\!]^{B})$ to learn $[\![ \epsilon ]\!]^{l}$.
    
    $P_0$ and $P_1$ compute and output $\lfloor \frac{x_0}{d} \rfloor + [\![ X_1 ]\!]^{l} + [\![ \epsilon ]\!]^{l}$.

\end{algorithm}

\vspace{5pt}
\noindent
\textbf{Correctness and security.}
The correctness of $\Pi_{\mathsf{\mathsf{Div}}}$ is ensured by Equation~\ref{equ:division faithful} and Equation~\ref{equ:epsilon}.
The security relies on the security of protocols for 
$\mathcal{F}_{\mathsf{MW}}$, 
$\mathcal{F}_{\mathsf{LUT}}$, 
$\mathcal{F}_{\mathsf{DReLU}}$, and
$\mathcal{F}_{\mathsf{B2A}}$.

\vspace{5pt}
\noindent
\textbf{Complexity.}
The two $\mathcal{F}_{\mathsf{LUT}}$ invoked by $\Pi_{\mathsf{\mathsf{Div}}}$ can be combined as one with $4$ entries, each with $l + l_d + 1$ bits.
Let $C_{\mathsf{MW}}$ denotes the communication of protocol for $\mathcal{F}_{\mathsf{MW}}$, the total communication of $\Pi_{\mathsf{\mathsf{Div}}}^{l,d}$ is $\lambda (l_d + 3) + 5l + 18 l_d + C_{\mathsf{MW}}$.


\subsubsection{Truncation protocol}

The truncation function can be viewed as a special case of the division function, where the divisor is a power of two, i.e., $d = 2^k$.
In this case, Equation~\ref{equ:division} can be rewritten as
$
        \mathsf{Trun}(x, k) 
        = \mathsf{Div}(x,2^k)
        = \lfloor \frac{x_0}{d} \rfloor + \lfloor \frac{x_1}{d} \rfloor - \mathsf{MW}(x) \cdot 2^{l - k} + e', 
$  
where $e' = \bm{1}\{x_0 \bmod 2^k + x_1 \bmod 2^k \geq 2^k\}$.
Accordingly, the truncation protocol can be implemented by invoking the computing $\mathsf{MW}(x)$ protocol, a $\mathsf{DReLU}$ (or comparison) protocol, and a $\mathsf{B2A}$ protocol.
The total communication of the truncation protocol is $\lambda(k + 1) + l + 13k + C_{\mathsf{MW}}$.



\subsection{Trigonometric Function}
\label{sec:app sin}

We consider the computation of the trigonometric function $\sin(x)$, where $\cos(x)$ can be computed using similar method. 
The function $\sin(x)$ fits the form of Equation~\ref{equ:func x y z} by applying the sum formula for sine, which expresses $\sin(a+b+c)$ as:
\begin{equation*}
    \begin{split}
        \sin(a+b+c) 
        &= \sin(a) \cos(b) \cos(c) 
        + \cos(a) \sin(b) \cos(c) \\
        &+ \cos(a) \cos(b) \sin(c) 
        - \sin(a) \sin(b) \sin(c).
    \end{split}
\end{equation*}
Therefore, the value of $\sin(\mathsf{Real}(x))$ can be computed using the following formulation:
\begin{equation}\label{equ:sin}
    \begin{split}
        \sin(\mathsf{Real}(x)) 
        &= \sin(\frac{x_0 + x_1 - \mathsf{MW}(x) \cdot L}{2^f}) \\
        &= 
        \sin(\frac{x_0}{2^f}) \cos(\frac{x_1}{2^f}) \cos(-\mathsf{MW}(x) \cdot \frac{L}{2^f})\\
        &+\cos(\frac{x_0}{2^f}) \sin(\frac{x_1}{2^f}) \cos(-\mathsf{MW}(x) \cdot \frac{L}{2^f})\\
        &+\cos(\frac{x_0}{2^f}) \cos(\frac{x_1}{2^f}) \sin(-\mathsf{MW}(x) \cdot \frac{L}{2^f})\\
        &-\sin(\frac{x_0}{2^f}) \sin(\frac{x_1}{2^f}) \sin(-\mathsf{MW}(x) \cdot \frac{L}{2^f}).
    \end{split}
\end{equation}
The computation of $\sin(\mathsf{Real}(x))$ involves one invocation of the $\mathsf{MW}(x)$, several multiplications, and two LUT protocols. 
The details are listed in Appendix~\ref{app:sin}.

\subsection{Exponential Function}
\label{sec:app exp}

The exponential function is a typical instance of the function in Equation~\ref{equ:func int}.
Let $a > 0$ be a public base and let $x$ be an exponent with bitwidth $l$ and precision $f$.
Then, $a^{\mathsf{Real}(x)}$ can be computed as: 
\begin{equation}\label{equ:exp}
    a^{\mathsf{Real}(x)} = a^{\frac{x_0}{2^f}} \cdot a^{\frac{x_1}{2^f}} \cdot a^{\frac{-\mathsf{MW}(x) \cdot L}{2^f}}.
\end{equation}
However, when $l$ is much larger than $f$, the bitwidth required to represent $a^{\frac{x_i}{2^f}}$ is large, increasing significantly overhead for multiplying $a^{\frac{x_0}{2^f}}$ and $a^{\frac{x_1}{2^f}}$.
To mitigate this issue, we set $l = f + \alpha$ and restrict the input range to $[-2^{\alpha-1}, 2^{\alpha-1})$.
Let $A_i = a^{\frac{x_i}{2^f}}$ for $i \in \{0,1\}$.
Then, the range of $A_i$ is $1$ to $a^{\frac{L}{2^f}}=a^{2^{\alpha}}$.
The integer part of $A_i$ can be represented using $\mu = \lceil \frac{L}{2^{f}} \log a \rceil + 1$ bits, where the additional one bit is used for the sign. 
We also assign a precision of $f_A$ bits to $A_0$ and $A_1$, resulting in a total bitwidth of $l_A = \mu + f_A$.
In practice, we set $\alpha = 3$ and $f_A = 10$, which achieves both high efficiency and accuracy.
For computing $a^{\frac{-\mathsf{MW}(x) \cdot L}{2^f}}$, which can take values of $1$, $a^{-2^{\alpha}}$ and $a^{-2^{\alpha+1}}$, a higher precision $f_M$ is required to ensure accuracy, since the latter two values are very close to zero.
Fortunately, $f_M$ has minimal impact on the total communication cost of the protocol, and can be set large.
For the setting $\alpha = 3$, we set $f_M = 32$, and use a total bitwidth of $l_M = f_M + 2$, including one bit for the integer part and one for the sign.

Based on these parameter settings, we present our exponential protocol $\Pi_{\mathsf{Exp}}^{l,f,l',f'}$ in Algorithm~\ref{alg:exponential}.
This protocol assumes the input $x$ is shared over $l= f + 3$ bits ring.
However, it can also be applied when $x$ is shared over a larger ring $\mathbb{Z}_{2^{l_r}}$, as long as $|x| < 2^{f+2}$.
In such case, given an input $[\![ x ]\!]^{l_r}$, we compute $[\![ x ]\!]^{f+3}_i = [\![ x ]\!]^{l_r}_i \mod 2^{f+3}$ for $i \in \{0,1\}$ to get $[\![ x ]\!]^{f+3}$, and use $[\![ x ]\!]^{f+3}$ as input to $\Pi_{\mathsf{Exp}}^{l,f,l',f'}$ to compute $a^{\mathsf{Real}(x)}$.





\begin{algorithm}[!htb]
	\caption{Computing $a^x$,
    $\Pi_{\mathsf{Exp}}^{l, f, l', f'}$: }
	\label{alg:exponential}
	\LinesNumbered 

	\KwIn{$P_0$ and $P_1$ hold $[\![ x ]\!]^{l}$ with precision $f$ where $l = f+3$, and a public positive $a$.}

	\KwOut{$P_0$ and $P_1$ output $[\![ Exp ]\!]^{l'}$ with precision $f'$, such that $\mathsf{Real}(Exp) = a^{\mathsf{Real}(x)}$.}


    Let $\mu = \lceil 2^{l-f} \cdot \log a \rceil + 1$, $f_A = 10$, and $l_A = \mu + f_A$.

    Let $f_M = 32 $ and $l_M= f_M + 2$.


    For $i \in \{0,1\}$, $P_i$ locally computes $A_i = a^{\frac{x_i}{2^{f}}}$, and encode $A_i$ as $\hat{A}_i=\mathsf{Fix}(A_i, l_A, f_A)$.

    $P_0$ and $P_1$ invoke $\mathcal{F}_{\mathsf{Mul}}^{l_A,l_A}(\hat{A}_0, \hat{A}_1)$ and learn $[\![ B ]\!]^{2l_A}$.

    \For{$j \in \{0,1,2\}$}
    {
        Parties computes $M_j = a^{-\frac{j \cdot L}{2^f}}$, and encodes it as $\hat{M}_j = \mathsf{Fix}(M_j, l_M, f_M)$.
    
        $P_0$ and $P_1$ invoke $\mathcal{F}_{\mathsf{Mul}}^{2l_A, l_M}([\![ B ]\!]^{2l_A}, \hat{M}_j)$ and learn $[\![Exp_j ]\!]^{2l_A+l_M}$.


        For $i \in \{0,1\}$, $P_i$ computes $[\![Exp_j ]\!]^{2\mu + 2 + f'}_i = [\![Exp_j ]\!]^{2l_A+l_M}_i \gg (2f_A + f_M - f') \bmod 2^{2\mu + 2 + f'}$.

        $P_0$ and $P_1$ set $[\![ T ]\!][j]^{2\mu + 2 + f'} = [\![Exp_j ]\!]^{2\mu + 2 + f'}$.
    }

    $P_0$ and $P_1$ invoke $\mathcal{F}_{{\mathsf{MW}}}^{l, 2}([\![ x ]\!]^{l})$ to learn $[\![ \mathsf{MW} ]\!]^{2}$.

    $P_0$ and $P_1$ invoke $\mathcal{F}_{\mathsf{LUT}}([\![T ]\!]^{2\mu + 2 + f'}, [\![ \mathsf{MW} ]\!]^{2})$ and learn $[\![Exp ]\!]^{2\mu + 2 + f'}$.

    For $i \in \{0,1\}$, $P_i$ computes $[\![Exp ]\!]_i^{l'} = [\![Exp ]\!]_i^{2\mu + 2 + f'} \mod 2^{l'}$.
    
    $P_0$ and $P_1$ output $[\![Exp ]\!]^{l'}$.
    
\end{algorithm}

\vspace{5pt}
\noindent
\textbf{Correctness and security.}
The correctness of $\Pi_{\mathsf{Exp}}$ is ensured by Equation~\ref{equ:exp}.
The security of $\Pi_{\mathsf{Exp}}$ relies on the security of protocols for 
$\mathcal{F}_{\mathsf{Mul}}$, 
$\mathcal{F}_{\mathsf{MW}}$, and
$\mathcal{F}_{\mathsf{LUT}}$. 

\vspace{5pt}
\noindent
\textbf{Complexity.}
Due to the properties of exponential function, both the $A_0$ and $A_1$ are positive, and the $\mathcal{F}_{\mathsf{Mul}}^{l_A, l_A}$ in line 4 can be implemented using $\mathcal{F}_{\mathsf{CrossTerm}}^{l_A, l_A}$.
Furthermore, the three instances of $\mathcal{F}_{\mathsf{Mul}}^{2l_A, f_M +2}$ (in line 7) can be realized by a single invocation of $\mathcal{F}_{\mathsf{SExt}}^{2l_A, 2l_A + f_M +2}$, as $\hat{M}_j$ is public and the value of $B$ is unchanged for $j\in \{0,1,2\}$.
Moreover, since $|\hat{A}_0 \cdot \hat{A}_1| < 2^{l_A - 2}$, the $\mathcal{F}_{\mathsf{SExt}}^{2l_A, 2l_A + f_M +2}$ can be efficiently implemented using the method proposed in \cite{geo}, which incurs small overhead.
Let $C_{\mathsf{MW}}$ denote the communication cost of $\mathcal{F}_{\mathsf{MW}}^{l,2}$.
The total communication cost of the exponential protocol is approximately:
$\lambda (l_A + 3) + \frac{3}{2}l_A^2 + 8\mu + 4f' + C_{\mathsf{MW}}$, where $\mu = \lceil 8 \cdot \log a \rceil + 1$ and $l_A = \mu + 10$.
For the case where $x$ is shared over a large ring $\mathbb{Z}_{2^{l_r}}$ for $l_r > f+3$, $\Pi_{\mathsf{MW_{conv}}}$ is invoked to compute $\mathsf{MW}$, with communication $2(\lambda + 2)$.
In this scenario, the total communication of $\Pi_{\mathsf{Exp}}$ is reduced to  
$\lambda (l_A + 5) + \frac{3}{2}l_A^2 + 8\mu + 4f'$.






\subsubsection{Evaluating $e^{-x}$}
\label{subsec:e^-x}

Considering the function $e^{-x}$, for $x \geq 0$, we first divide the range of $x$ into two intervals: $x \in [0,8) \bigcup [8, +\infty)$.
When $x$ belongs to the second interval, we have $e^{-x} \leq e^{-8} \leq 3.35 \times 10^{-4}$, which is negligible. 
Consequently, $e^{-x}$ is approximated as $0$ for $x \geq 8$, resulting in the ULP error less than $1.37$.
For $x \in [0,8)$, let $y = -x + 4$, then $e^{-x} = e^y \cdot e^{-4}$ and $y \in (-4,4]$.
However, the input range of $\Pi_{\mathsf{Exp}}$ is $[-4,4)$.
To address this discrepancy, we set $y'= y-\tau$, where $\tau = 2^{-f}$.
Then, $e^{-x}$ is approximated as $e^{y'} \cdot e^{-4}$, and $y' \in [-4, 4)$ can serve as input of $\Pi_{\mathsf{Exp}}$.
The details of our evaluating $e^{-x}$ protocol are shown in Algorithm~\ref{alg:r exponential}.
In this algorithm, the $\mathsf{DReLU}$ protocol is employed to determine whether $x$ lies in interval $[0,8)$ or $[8, +\infty)$.
For $x \in [0,8)$, $\Pi_{\mathsf{Exp}}$ is utilized to compute $e^{-x}$.
Moreover, if $l = f + 4$, then the range of $x$ is $[-8, 8)$.
In this case, it is unnecessary to check whether $x > 8$, allowing the $\mathsf{DReLU}$ protocol to be omitted, and only a single invocation of $\Pi_{\mathsf{Exp}}$ is required.


\begin{algorithm}[!htb]
	\caption{Computing $e^{-x}$ for $x \geq 0$, $\Pi_{\mathsf{rExp}}^{l, f}$: }
	\label{alg:r exponential}
	\LinesNumbered 

	\KwIn{$P_0$ and $P_1$ hold $[\![ x ]\!]^{l}$ with precision $f$, where $\mathsf{int}(x) \geq 0$ and $l \geq f + 4$.}
 
	\KwOut{$P_0$ and $P_1$ output $[\![ rExp ]\!]^{l}$ with precision $f$, such that $\mathsf{Real}(rExp) = e^{-\mathsf{Real}(x)}$.}


    $P_0$ and $P_1$ compute $[\![ z ]\!]^{f+3} = -[\![ x ]\!]^{l} + 4 \cdot 2^{f} - 1 \mod 2^{f+3}$.


    $P_0$ and $P_1$ invoke $\mathcal{F}_{\mathsf{Exp}}^{f+3, f, l, f}([\![ z ]\!]_i^{f+3})$ to learn $[\![ sExp ]\!]^{l'}$.

    $P_0$ and $P_1$ compute $[\![ Exp ]\!]^{l} = [\![ sExp ]\!]^{l} \cdot \mathsf{Fix}(e^{-4}, l, f)$.

    $P_0$ and $P_1$ invoke $\mathcal{F}_{\mathsf{DReLU}}^{l}(-[\![ x ]\!]^{l}+8\cdot2^{f})$ to learn $[\![ b ]\!]^{B}$.

    $P_0$ and $P_1$ invoke $\mathcal{F}_{\mathsf{MUX}}([\![ Exp ]\!]^{l}, [\![ b ]\!]^{B})$ to learn $[\![ rExp ]\!]^{l}$.

    $P_0$ and $P_1$ output $[\![ rExp ]\!]^{l}$.

\end{algorithm}

\vspace{5pt}
\noindent
\textbf{Correctness and security.}
The correctness of $\Pi_{\mathsf{rExp}}$ comes from that for $x \in [0,8)$, we have $e^{-x} \approx e^{-x+4-2^{-f}} \cdot e^{-4}$.
The security relies on the security of protocols for 
$\mathcal{F}_{\mathsf{Exp}}$,
$\mathcal{F}_{\mathsf{DReLU}}$, and 
$\mathcal{F}_{\mathsf{MUX}}$.

\vspace{5pt}
\noindent
\textbf{Complexity.}
$\Pi_{\mathsf{rExp}}^{l, f}$ needs one call for each of 
$\mathcal{F}_{\mathsf{DReLU}}^{l}$, 
$\mathcal{F}_{\mathsf{Exp}}^{l,f, l, f}$, and 
$\mathcal{F}_{\mathsf{MUX}}^{l}$.
Since $l > f + 3$, the $\mathcal{F}_{\mathsf{Exp}}^{l,f, l, f}$ invokes $\Pi_{\mathsf{MW_{conv}}}$ to compute $\mathsf{MW}$.
Moreover, in $\mathcal{F}_{\mathsf{Exp}}^{l,f, l, f}$, $\mu = 13$ and $l_A = 23$.
Therefore, the total communication of $\Pi_{\mathsf{rExp}}^{l, f}$ is approximately 
$\lambda (l + 29) + 18l + 4f  + 897$.
Moreover, if $l = f + 4$, then the overhead of $\Pi_{\mathsf{rExp}}^{l, f}$ is the same as $\Pi_{\mathsf{Exp}}^{l, f,l,f}$, approximated as
$28 \lambda  + 4f + 2l + 897$.

\subsubsection{Evaluating $\mathsf{Softmax}$}

The $\mathsf{Softmax}$ function is a commonly used activation function in machine learning, particularly in multi-class classification tasks.
For an input vector $\vec{z} = \{z_0, z_1,...,z_{n-1}\}$, $\mathsf{Softmax}$ is defined as:
\begin{equation}
    \mathsf{Softmax}(z_i) = \frac{e^{z_i}}{\sum_{j=0}^{n-1} e^{z_j}}.
\end{equation}
In both PPML and plaintext implementations, each $z_i$ is normalized as $z_i^* = z_i - z_{max}$, where $z_{max}$ is the maximum value in $\vec{z}$.
This normalization ensures that the inputs to the exponential function are non-positive while preserving the correctness of $\mathsf{Softmax}$.
Consequently, the protocol $\Pi_{\mathsf{rExp}}$ can be used to compute $e^{z_i^*}$.
The division required for $\mathsf{Softmax}$ can be implemented using SirNN's method \cite{SiRnn}. 
First, the reciprocal $D = \frac{1}{\sum_{i=0}^{n-1} e^{z_i}}$ is computed.
Then $\mathsf{Softmax}(z_i)$ is obtained by computing $e^{z_i} \cdot D$ for $i \in \{0,1,...,n-1\}$.
Since $D$ is computed only once for the entire vector, its amortized overhead is negligible.
In the multiplication step, the same $D$ is multiplied by $n$ different values. 
Therefore, these multiplications are reformulated as a matrix multiplication $[D] \cdot [z_0, z_1, ..., z_{n-1}]$, and computed by invoking the matrix multiplication protocol in SirNN \cite{SiRnn}.
Finally, we can get $\mathsf{Softmax}(z_i)$ for $i \in \{0,1,...,n-1\}$.

%% file: sec/experiment_NEW.tex
\section{Experiments}
\label{sec:experiment}

\textbf{Experimental setup.}
Our experiments were conducted on a server equipped with Intel(R) Xeon(R) Platinum processors operating at 2.5 GHz, featuring 8 logical CPUs and 16 GB of memory.
To simulate various network conditions, we used Linux Traffic Control (tc), emulating a LAN environment with 1 Gbps bandwidth and 0.8 ms RTT latency, and a WAN environment with 100 Mbps bandwidth and 80 ms RTT latency.
All experiments were built upon the SCI library~\cite{SCI}, with the underlying IKNP-style OT protocols.
Our implementations are publicly available at \url{https://anonymous.4open.science/r/usenix2026}.

\subsection{Computing $\mathsf{MW}(x)$}


Prior work \cite{geo} proposed an efficient method for computing $\mathsf{MW}(x)$ under the constraint $|x| < \frac{L}{3}$.
In this work, we extend this approach by relaxing the upper bound to $|x| < B$ for any $B \leq \frac{L}{2}$.
We conduct experiments on our computing $\mathsf{MW}(x)$ protocol $\Pi_{\mathsf{MW}}$, under varying constraints of $|x| < B$, where $B \in \{0.5 \cdot \frac{L}{2}, 0.8 \cdot \frac{L}{2}, 0.9999 \cdot \frac{L}{2}, 0.999999 \cdot \frac{L}{2}, \frac{L}{2}\}$.
These settings demonstrate the performance of $\Pi_{\mathsf{MW}}$ across different bounds.
The experimental results are shown in Table~\ref{tab:exp MW}.
For $B = 0.5 \cdot \frac{L}{2}$, our $\Pi_{\mathsf{MW}}$ is the same as the protocol in \cite{geo}, achieving maximum efficiency.
However, for $B > \frac{L}{3}$, the method in \cite{geo} fails, while our $\Pi_{\mathsf{MW}}$ remains effective.
If it is known that $|x| < 0.9999 \cdot \frac{L}{2}$, which makes $99.99\%$ of the values in $\mathbb{Z}_{L}$ valid, $\Pi_{\mathsf{MW}}$ achieves approximately a $2.5 \times$ improvement in efficiency compared to the baseline.
Even for $B = 0.999999 \cdot \frac{L}{2}$, where only $0.0001\%$ of the values on $\mathbb{Z}_{L}$ are invalid, $\Pi_{\mathsf{MW}}$ still provides nearly a $2 \times$ efficiency improvement over the baseline method.
This implies that the efficiency of $\Pi_{\mathsf{MW}}$ can be significantly improved even there is a very small gap between $B$ and $\frac{L}{2}$.
For the case there is no priori knowledge on the range of $x$, the baseline method is used, which invoke only one $l$-bit comparison protocol.

\begin{table}[!htb]
    \centering
    \caption{Communication and runtime for protocol $\Pi_{\mathsf{MW}}^{l,l'}$, where $l = 37$, $l' = 2$ and $L = 2^l$.
    The communication and runtime are accumulated for $2^{20}$ runs of the protocols.
    }
    \label{tab:exp MW}
    \setlength{\tabcolsep}{10pt}
 \begin{tabular}{cc ccc}
    \toprule

    \multirow{2}*{$B$} 
    & \multicolumn{2}{c}{Time (s)} 
    & \multirow{2}*{Comm. (MB)} 
    \\

    \cline{2-3}
    & LAN & WAN & \\
    \midrule

    $0.5 \cdot \frac{L}{2}$ &0.19  & 2.04 & 16.25\\

    
    $0.8 \cdot \frac{L}{2}$ & 0.67 &5.51  & 49.25 \\
    
    $0.9999 \cdot \frac{L}{2}$ &4.94  & 28.19  & 262.25 \\
    
    $0.999999 \cdot \frac{L}{2}$ &6.35  &35.65   & 336.25 \\
    
    $1 \cdot \frac{L}{2}$ (baseline) &12.13  &71.08  & 671.75  \\

    \bottomrule
	\end{tabular}
    
\end{table}

\subsection{Experiments on Real-world Functions}
\label{sec:exp app}



\begin{table}
    \centering
    \caption{Comparing the runtime and communication of division protocol in CrypTFlow2 \cite{CrypTFlow2} with ours.
    For $x \in \mathbb{Z}_L$, we consider constraint $B \in \{B_1, B_2, B_3, \frac{L}{2}\}$, where $B_1 = 0.5 \cdot \frac{L}{2}$, $B_2 = 0.99 \cdot \frac{L}{2}$ and $B_3 = 0.999999 \cdot \frac{L}{2}$.
    The communication and runtime are accumulated for $2^{18}$ runs of the protocols.
    }
    \label{tab:exp div}
    \setlength{\tabcolsep}{9pt}
 \begin{tabular}{ccc ccc}
    \toprule

    \multirow{2}*{Method} 
    & \multirow{2}*{$B$} 
    & \multicolumn{2}{c}{Time (s)} 
    & \multirow{2}*{Comm. (MB)} 
    \\

    \cline{3-4}
    & & LAN & WAN & \\
    \midrule

     \cite{CrypTFlow2}
    & -
    & 7.75
    & 48.24
    & 354.72
    \\
    
\hline

    \multirow{8}*{\makecell{Ours}}
    & \multirow{2}*{$B_1$}
    & 1.60
    & 11.36
    & 82.53
    \\
    \cline{3-5}
    & & $\mathbf{4.84\times}$
    & $\mathbf{4.24 \times}$
    & $\mathbf{4.29 \times}$
    \\

\cline{2-5}
    & \multirow{2}*{$B_2$}
    & 2.12
    & 15.28
    & 113.09
    \\
    \cline{3-5}
    & & $\mathbf{3.65 \times}$
    & $\mathbf{3.15 \times}$
    & $\mathbf{3.13 \times}$
    \\

\cline{2-5}
    & \multirow{2}*{$B_3$}
    & 3.84
    & 22.26
    & 167.59
    \\
    \cline{3-5}
    & & $\mathbf{2.01 \times}$
    & $\mathbf{2.16 \times}$
    & $\mathbf{2.11 \times}$
    \\
    
\cline{2-5}
    & \multirow{2}*{$\frac{L}{2}$}
    &  5.46
    &  32.99
    &  252.72
    \\
    \cline{3-5}
    & & $\mathbf{1.41 \times}$
    & $\mathbf{1.46 \times}$
    & $\mathbf{1.40 \times}$
    \\

    \bottomrule
	\end{tabular}
    
\end{table}

\begin{table}[!htb]
    \centering
	\caption{
    Communication and runtime for our evaluating $\sin(x)$ protocol $\Pi_{\mathsf{sin}}$.
    The bitwidth and precision are $l = 21$ and $f = 12$.
    For $x \in \mathbb{Z}_L$, the values of constraint $B$ are set as $B_1 = 0.5 \cdot \frac{L}{2}$, $B_2 = 0.99 \cdot \frac{L}{2}$, $B_3 = 0.999999 \cdot \frac{L}{2}$ and $\frac{L}{2}$.
    Symbols "aULP" and "mULP" denote the average and maximum ULP error, respectively.
    The communication (in MB) and runtime are accumulated for $2^{18}$ runs of the protocols.
    }
	\label{tab:exp sin}
    \setlength{\tabcolsep}{9pt}{
 \begin{tabular}{cccccc}
    \hline
    
    \multirow{2}*{$B$}   & \multicolumn{2}{c}{Time. (s)} &  \multirow{2}*{Comm.} & \multirow{2}*{aULP} & \multirow{2}*{mULP}\\
     \cline{2-3}
      & LAN & WAN &  &  \\

    \hline
    
    $B_1$  &4.73 &43.44 & 381.59&0.500 &1.280 \\
    
    $B_2$  &5.52 & 46.64&407.84 & 0.505 &1.295 \\

    $B_3$  &6.77 & 55.06& 461.09&0.504 & 1.304\\

    $\frac{L}{2}$  &7.07 &55.76 &474.46 & 0.504 & 1.308\\
    
    \hline

	\end{tabular}}
\end{table}

\begin{table*}[htb]
    \centering
	\caption{
    Comparing the runtime and communication costs of our evaluating $e^{-x}$ protocol with SirNN~\cite{SiRnn} and Bolt~\cite{bolt}.
    For SirNN, the bitwidth is set as $l = 16$, and in Bolt, $l = 37$.
    All the precision is set as $f = 12$.
    The communication and runtime are accumulated for $2^{18}$ runs of the protocols.}
	\label{tab:exp e**x}
    \setlength{\tabcolsep}{19pt}{
 \begin{tabular}{ccccccc}
    \hline
    
    \multirow{2}*{$l$} & \multirow{2}*{Method.}  & \multicolumn{2}{c}{Time. (s)} &  \multirow{2}*{Comm. (MB)} & \multirow{2}*{avg ULP} & \multirow{2}*{max ULP} 
    \\
     \cline{3-4}
     & & LAN & WAN & 
     \\

    \hline

     \multirow{2}*{16} & SirNN  \cite{SiRnn}  & 10.30 & 50.49 &501.12 & 1.183& 2.640 
     \\

     & Ours  & 1.86& 17.75&  156.90 & 0.353 & 1.435 
     \\

    \cline{3-5}
      & & $\mathbf{5.53 \times}$ & $\mathbf{2.84 \times}$ & $\mathbf{3.19 \times}$ \\

    \hline
    
    \multirow{2}*{37} & Bolt \cite{bolt} &15.09 & 98.22& 944.28  & 0.010& 8.681 
    \\

    
    & Ours  &  4.88& 34.58&297.81  & 0.004 & 1.435 
    \\

     \cline{3-5}
      & & $\mathbf{3.09 \times}$ & $\mathbf{2.84 \times}$ & $\mathbf{3.17 \times}$ \\
    
    \hline

	\end{tabular}}
\end{table*}

\subsubsection{Division protocol}

For $x \in \mathbb{Z}_{L}$, we perform four experiments on our division protocol $\Pi_{\mathsf{Div}}$, with the constraints $|x| < B$ for $B \in \{0.5 \cdot \frac{L}{2}, 0.9999 \cdot \frac{L}{2}, 0.999999 \cdot \frac{L}{2}, \frac{L}{2}\}$ and compare the performance with CrypTFlow2's division protocol \cite{CrypTFlow2}.
The experimental results are presented in Table~\ref{tab:exp div}.
When $B = 0.5 \cdot \frac{L}{2}$, $\Pi_{\mathsf{Div}}$ achieves the highest efficiency, demonstrating an improvement of $4.24 \times$ to $4.84 \times$ compared to CrypTFlow2's method. 
For $B = 0.999999 \cdot \frac{L}{2}$, our protocol outperforms CrypTFlow2 by $2.01 \times$ to $2.16 \times$.
Even for the case there is no extra constraint on $|x|$, $\Pi_{\mathsf{Div}}$ achieves an approximately $1.4 \times$ improvement.
These experimental results demonstrate that the performance gains of our $\Pi_{\mathsf{Div}}$ are not only from the optimization of computing $\mathsf{MW}(x)$ protocol, but also from the novel design of the division protocol.
Moreover, $\Pi_{\mathsf{Div}}$ produces the exact result $\lfloor \frac{x}{d} \rfloor$ without error.

\subsubsection{$\sin(x)$ protocol}

We apply experiments on our trigonometric protocol $\Pi_{\mathsf{sin}}$, with the experimental results presented in Table~\ref{tab:exp sin}.
The communication increases only slightly with the constraint $B$, as the primary overhead arises from the multiplication protocols.
The experimental results also demonstrate the high accuracy of $\Pi_{\mathsf{sin}}$, with maximum ULP error as approximately $1.3$.
Variations in the ULP error are attributed to the randomness of the input values in each experimental group.

    

    
    


    


\subsubsection{Exponential protocol}

The experimental results for our exponential protocol $\Pi_{\mathsf{rExp}}$ are presented in Table~\ref{tab:exp e**x}, with all precisions set as $f = 12$.
When compared with SirNN~\cite{SiRnn}, we adopted SirNN's settings by choosing $l = 16$, and $l = f+4$.
Under these settings, the input range of $e^{-x}$ is $x \in [0,8)$.
Consequently, the $\mathsf{DReLU}$ protocol in $\Pi_{\mathsf{rExp}}$ can be removed.
Our $\Pi_{\mathsf{rExp}}$ achieves a $2.84 \times$ to $5.53 \times$ improvement in efficiency compared to SirNN. 
For accuracy, although SirNN achieves high accuracy for evaluating $e^{-x}$, our method achieves even higher accuracy, with a maximum ULP error of $1.435$ at a precision of $f=12$, corresponding to an error of approximately $3 \cdot 10^{-4}$ in floating-point.
The ULP error is computed by testing all the inputs in interval $[0,8)$.
When comparing with Bolt~\cite{bolt}, we adjusted the bitwidth to $l = 37$.
In this configuration, the $\mathsf{DReLU}$ protocol is invoked, incurring greater overhead in $\Pi_{\mathsf{rExp}}$ compared to the $l = 16$ case.
Despite this, $\Pi_{\mathsf{rExp}}$ achieves a $2.84 \times$ to $3.17 \times$ efficiency improvement over Bolt.
In terms of accuracy, we compute ULP error for $e^{-x}$ with range $x \in (0,1000]$.
The maximum ULP error for our method is $1.435$, significantly outperforming prior works.

\subsubsection{$\mathsf{Softmax}$ protocol}

For $\mathsf{Softmax}$ function, the input used in our experiments is a $128 \times 768$ matrix, where $128$ $\mathsf{Softmax}$ functions are performed, each with input length as $768$.
This setup reflects the typical parameters of the $\mathsf{Softmax}$ function employed in BERT (Bidirectional Encoder Representations from Transformers)~\cite{bert}.
The experimental results are listed in Table~\ref{tab:exp softmax}.
The performance improvements of our $\mathsf{Softmax}$ protocol stem from two key optimizations: our novel protocol for evaluating $e^{-x}$, and an optimized batch division protocol. 
These enhancements enable a $4.6 \times$ to $6.2 \times$ improvement compared to Iron~\cite{iron}.
When compared with Bolt~\cite{bolt}, our protocol reduces communication costs to $42\%$ of its original level, and achieves an approximately $2 \times$ improvement in runtime.

\begin{table}[htb]
    \centering
    \caption{
    Comparing the runtime and communication costs of our $\mathsf{Softmax}$ protocol with Bolt \cite{bolt} and Iron \cite{iron}.
    The bitwidth and precision are set as $l = 37$ and $f = 12$.
    The input of $\mathsf{Softmax}$ protocol is a $128 \times 768$-dimensional matrix, where there are $128$ vectors, each with dimension $768$.
    }
    \label{tab:exp softmax}
    \setlength{\tabcolsep}{14pt}
 \begin{tabular}{cc ccc}
    \toprule

    \multirow{2}*{Method} 
    & \multicolumn{2}{c}{Time (s)} 
    & \multirow{2}*{Comm. (MB)} 
    \\

    \cline{2-3}
    & LAN & WAN & \\
    \midrule
     Iron \cite{iron} & 29.67 & 174.11 & 1660.68   \\
     Bolt \cite{bolt} & 9.80 & 73.63 & 630.16   \\
     Ours & 4.78 & 37.91 & 270.84   \\
    \bottomrule
	\end{tabular}
    
\end{table}

%% file: sec/Conclusion.tex
\section{Conclusion}

This work proposes a new method to design secure two-party protocols for a class of real-world functions with real-number inputs.
We first generalize the computing $\mathsf{MW}(x)$ protocol to accommodate arbitrary constraint. 
Building on this foundation, we give an efficient method for computing signed real number from shares.
Then we propose secure evaluation methods for a variety of real-world functions, including division, trigonometric, and exponential functions. Our approach enables these functions to be evaluated with both low overhead and high accuracy.
This work introduces a novel perspective on protocol design, and we anticipate that the ideas presented here will inspire broader advancements in protocol development and optimization.

%% file: App/proofs.tex
\section{Proofs}
\label{app:proof}

\subsection{Proof of Lemma~\ref{lem:comp A}}
\label{app:proof lem:comp A}

\begin{proof}

    To prove this lemma, we need to prove that $x > y$ if and only if $\lfloor \frac{x}{A} \rfloor > \lfloor \frac{y}{A} \rfloor$.
    For necessity, $x > y$ if and only if $x \geq y + A$ as $x - y \in [A, L) \bigcup [-L, 0)$.
    Therefore $\frac{x}{A} \geq \frac{y}{A} + 1$, then we have $\lfloor \frac{x}{A} \rfloor > \lfloor \frac{y}{A} \rfloor$.

    For sufficiency, we use the method of proof by contradiction.
    Suppose $x \leq y$, we have $\frac{x}{A} \leq \frac{y}{A}$ and $\lfloor \frac{x}{A} \rfloor \leq \lfloor \frac{y}{A} \rfloor$, which contradicts the condition $\lfloor \frac{x}{A} \rfloor > \lfloor \frac{y}{A} \rfloor$.
    Therefore, we have $x > y$.
\end{proof}

\subsection{Proof of Lemma~\ref{lem:wrap A}}
\label{app:proof lem:wrap A}

\begin{proof}
    As $x_0 + x_1 \neq L$, we have that $\mathsf{Wrap}(x_0, x_1, L) = \bm{1}\{x_0 + x_1 \geq L\} = \bm{1}\{x_0 + x_1 < L\} \oplus 1 = \mathsf{Comp}(x_1, L-x_0) \oplus 1$.
    As $A \neq 0$, we have $\mathsf{Comp}(x_1, L-x_0) \oplus 1 = \mathsf{Comp}(L-x_0, x_1) $.
    Let $a = x_1$ and $b = L - x_0$, we have $a - b = x_0 + x_1 - L \in [-L,0)\bigcup [A, L)$.
    Then from Lemma~\ref{lem:comp A} we can have that $\mathsf{Comp}(b, a) = \mathsf{Comp}(\lfloor \frac{b}{A} \rfloor, \lfloor \frac{a}{A} \rfloor)$.
    Therefore, $\mathsf{Wrap}(x_0, x_1, L) = \mathsf{Comp}( L-x_0, x_1) = \mathsf{Comp}(\lfloor \frac{ L-x_0}{A} \rfloor, \lfloor \frac{x_1}{A} \rfloor)$.
\end{proof}

\subsection{Proof of Theorem~\ref{theo:MW}}
\label{app:proof theo:MW}

\begin{proof}
    For $|x| < B$, we have $x_0 + x_1 \in [0,B) \bigcup [L - B, L + B) \bigcup [2L - B, 2L)$.
    The theorem is proven by considering the following cases:
    \begin{itemize}
        \item 
        For $x_0 + x_1 \in [0,B)$, $\mathsf{MW}(x) = 0$.
        In this case we have $x_0 \in [0,B)$, and thus $\Delta = 0$ and $x_0^* = x_0 - B + L$.
        Then $x_0^* + x_1 = x_0 + x_1 + L - B \in [L - B, L)$, which means $x_0^* + x_1 < 2L - 2B$ as $B < \frac{L}{2}$ and $M^* = 0$.
        Therefore, $\mathsf{MW}(x) = M^* + \Delta = 0$.

        \item 
        For $x_0 + x_1 \in [L - B, L +B)$, $\mathsf{MW}(x) = 1$.
        \begin{itemize}
            \item 
            If $x_0 < B$, we have $\Delta = 0$ and $x_0^* = x_0 - B + L$.
            Then $x_0^* + x_1 = x_0 + x_1 + L - B \in [2L - 2B, 2L)$, and thus $M^* = 1$.

            \item 
            If $x_0 > B$, we have $\Delta = 1$ and $x_0^* = x_0 - B$.
            Then $x_0^* + x_1 = x_0 + x_1 - B \in [L - 2B, L)$, and thus $M^* = 0$.
        \end{itemize}
        Therefore, $\mathsf{MW}(x) = M^* + \Delta = 1$.
        
        \item 
        For $x_0 + x_1 \in [2L - B, 2L)$, $\mathsf{MW}(x) = 2$.
        In this case $x_0 > L - B$, and thus $\Delta = 1$ as $B < \frac{L}{2}$.
        Then, $x_0^* + x_1 = x_0 + x_1 - B \in [2L - 2B, 2L - B)$, which means $M^* = 1$.
        Therefore, $\mathsf{MW}(x) = M^* + \Delta = 2$.
    \end{itemize}

    In all cases, $\mathsf{MW}(x) = M^* + \Delta$, completing the proof.

\end{proof}

\subsection{Proof of Lemma~\ref{lem:MW conv 1}}
\label{app:proof lem:MW conv 1}

\begin{proof}\label{proof:lem:MW conv 1}
We study the relationship between $\mathsf{MW}(z, 2^l)$ and $\mathsf{MW}(y, 2^{l+1})$ in the following cases.
\begin{itemize}
    \item 
    \textbf{Case 1:} {$y_0 + y_1 \in [0, 2^{l-1})$} (point $P \in \mathcal{A}$).
    For $i \in \{0,1\}$, we have $y_i < 2^{l-1}$ and $z_i = y_i$.
    Therefore, $z_0 + z_1 = y_0 + y_1$ belongs to interval $[0, 2^{l - 1})$,  and $\mathsf{MW}(z, 2^l) = \mathsf{MW}(y, 2^{l+1}) = 0$.
    
    \item 
    \textbf{Case 2:} {$y_0 + y_1 \in [2^{l+1} - 2^{l-1}, 2^{l+1} + 2^{l-1})$}.
    In this case, we consider the following sub-cases:
    \begin{itemize}
        \item 
        \textbf{Case (i):} $y_0 < 2^l$ and $y_1 < 2^l$ (point $P \in \mathcal{B}_0$).
        For $i \in \{0,1\}$, $z_i = y_i$.
        Therefore, $z_0 + z_1 = y_0 + y_1$, belongs to interval $[2^{l+1} - 2^{l-1}, 2^{l})$.
        In this case, $b = 1$, $\mathsf{MW}(y, 2^{l+1}) = 1$, and $\mathsf{MW}(z, 2^l) = 2$.
        Thus, $\mathsf{MW}(z, 2^l) = \mathsf{MW}(y, 2^{l+1}) + b$.

        \item 
        \textbf{Case (ii):} $y_0 \geq 2^l$ and $y_1 \geq 2^l$ (point $P \in \mathcal{C}_0$).
        For $i \in \{0,1\}$, $z_i = y_i - 2^l$.
        Therefore, $z_0 + z_1 = y_0 + y_1 - 2^{l+1}$, belongs to interval $[0, 2^{l-1})$.
        In this case, $a = 1$, $\mathsf{MW}(y, 2^{l+1}) = 1$, and $\mathsf{MW}(z, 2^l) = 0$.
        Thus, $\mathsf{MW}(z, 2^l) = \mathsf{MW}(y, 2^{l+1}) - a$.

        \item 
        \textbf{Case (iii):} $y_0 \geq 2^{l}$ and $y_1 < 2^l$, or $y_0 < 2^l$ and $y_1 \geq 2^l$.
        In this case, $z_0 + z_1 = y_0 + y_1 - 2^l$, belongs to interval $2^l - 2^{l - 1}, 2^{l} + 2^{l - 1}$.
        Therefore, $\mathsf{MW}(z, 2^l) = \mathsf{MW}(y, 2^{l+1}) = 1$.
        
    \end{itemize}

    \item 
    \textbf{Case 3:} {$y_0 + y_1 \in [2^{l+2}-2^{l-1}, 2^{l+2})$} (point $P \in \mathcal{D}$).
    For $i \in \{0,1\}$, we have $y_i \geq 2^{l+1} - 2^{l-1}$, and $z_i = y_i \mod 2^l = y_i - 2^l$.
    Therefore, $z_0 + z_1 = y_0 + y_1 - 2^{l+1}$ belongs to interval $2^{l+1} - 2^{l-1}, 2^{l+1}$, which means $\mathsf{MW}(z,2^l) = \mathsf{MW}(y,2^{l+1}) = 2$.
    \end{itemize}  
    Note that both $a$ and $b$ are zeros except for case (i) and case (ii).
    Therefore, we have $\mathsf{MW}(z,2^l) = \mathsf{MW}(y,2^{l+1}) - a + b$.
\end{proof}

\subsection{Proof of Theorem~\ref{lem:MW conv 2}}
\label{app:proof lem:MW conv 2}

\begin{proof}\label{proof:lem:MW conv 2}
    Let $\hat{y} = \hat{y}_0 + \hat{y}_1$.
    The range of $\hat{y}_0 + \hat{y}_1$ is $\hat{y}_0 + \hat{y}_1 \in [0, 2^{l-1}) \bigcup [2^{l} - 2^{l-1}, 2^{l} + 2^{l -1}) \bigcup [2^{l+1} - 2^{l-1}, 2^{l+1})$, corresponding to $\mathsf{MW}(\hat{y}, 2^{l+1})$ takes value of $0$, $1$ and $2$, respectively.
    Moreover, $\hat{y}_0 + \hat{y}_1 \in [0, 2^{l-1})$ if and only if $b = 1$, and $\hat{y}_0 + \hat{y}_1 \in [2^{l+1} - 2^{l-1}, 2^{l+1})$ if and only if $a = 1$.
    Therefore, $a - b$ can be written as:
    \begin{equation*}\label{equ: a-b}
        a - b =
        \begin{cases}
            1, \quad &\mathsf{MW}(\hat{y}, 2^{l+1}) = 0 \\
            0, \quad &\mathsf{MW}(\hat{y}, 2^{l+1}) = 1 \\
            -1, \quad &\mathsf{MW}(\hat{y}, 2^{l+1}) = 2 \\
        \end{cases}, 
    \end{equation*}
    which can be summarized as $a - b = 1 - \mathsf{MW}(\hat{y}, 2^{l+1})$.
    We use the idea in Theorem~\ref{theo:MW} to compute $\mathsf{MW}(\hat{y})$, with the constraint $|y| < 2^{l}$ for $y \in \mathbb{Z}_{2^{l + 1}}$.
    Defining $\hat{M}^* = \bm{1}\{\hat{y}_0^* \geq 2^{l}\} \land \bm{1}\{\hat{y}_1 \geq 2^{l}\}$ and $\delta = \bm{1}\{\hat{y}_0 \geq 2^{l - 1}\}$, we have $\hat{M}^* = \bm{1}\{\hat{y}^*_0 + y_1 \geq 2^{l+1}\} = \bm{1}\{\hat{y}_0^* \geq 2^{l}\} \land \bm{1}\{\hat{y}_1 \geq 2^{l}\}$ and $\mathsf{MW}(\hat{y}) = \hat{M}^* + \delta$.
    Therefore, $a - b = 1 - \mathsf{MW}(\hat{y}) = 1 - \delta - \hat{M}^*$.
\end{proof}

\section{Evaluating $\sin(x)$ protocol}
\label{app:sin}

Our evaluating $\sin(x)$ protocol $\Pi_{\sin}$ is shown in Algorithm~\ref{alg:sin}.
In the multiplication protocol used in lines 4–7, the inputs are held by $P_0$ and $P_1$ locally.
Therefore, the method in subsection~\ref{subsec:mul} is used.
Moreover, we can use the known $\mathsf{MSB}$ method to improve the efficiently.
For instance, when computing the product $s_0 \cdot c_1$, we first set $\hat{s}_0 = s_0 + 2^f$ and $\hat{c}_1 = c_1 + 2^f$, ensuring both values are non-negative. 
Then, the product $\hat{s}_0 \cdot\hat{c}_1$ can be computed using a single invocation of $\mathcal{F}_{\mathsf{CrossTerm}}$.
The original product $s_0 \cdot c_1$ can be recovered via $s_0 \cdot c_1 = \hat{s}_0 \cdot\hat{c}_1 - 2^f \cdot c_1 - 2^f \cdot s_0 -2^{2f}$, where the last three terms can be computed locally.
When computing $(sc + cs) \cdot C_j$ and $(cc - ss) \cdot S_j$ for $j \in \{0,1,2\}$, the signed extension protocols $\mathcal{F}_{\mathsf{SExt}}$ are invoked since $C_j$ and $S_j$ can be seen as public constants (see Section~\ref{subsec:mul} for details).
Only two $\mathcal{F}_{\mathsf{SExt}}$ are required in the loop, one for $sc + cs$ and one for $cc - ss$, as these values are unchanged.
Moreover, as the ranges of $(sc + cs)$ and $(cc - ss)$ are small, $\mathcal{F}_{\mathsf{SExt}}$ can be implemented using the method in \cite{geo}, resulting in very small overhead.
Finally, after retrieving the value of $\sin(x)$ from the lookup table, an additional $\mathcal{F}_{\mathsf{SExt}}$ is invoked to extend the bitwidth to $l'$, which can also be implemented using the method from \cite{geo}.

\begin{algorithm}[htb]
	\caption{Computing $\sin(x)$, $\Pi_{\mathsf{sin}}^{l, f, l', f'}$: }
	\label{alg:sin}
	\LinesNumbered 

	\KwIn{$P_0$ and $P_1$ hold $[\![ x ]\!]^{l}$ with precision $f$.}
 
	\KwOut{$P_0$ and $P_1$ output $[\![ z ]\!]^{l'}$ with precision $f'$, such that $\mathsf{Real}(z) = \sin(\mathsf{Real}(x))$.}

    Let $f_t = 14$, $l_t = f_t + 2$, $f_T = 30$ and $l_T = f_T + 2$.

    For $i \in \{0,1\}$, $P_i$ computes $\sin(\frac{x_i}{2^f})$ and $\cos(\frac{x_i}{2^f})$, and encode them as $s_i = \mathsf{Fix}(\sin(\frac{x_i}{2^f}), l_t, f_t)$ and $c_i = \mathsf{Fix}(\cos(\frac{x_i}{2^f}), l_t, f_t)$.


     $P_0$ and $P_1$ invoke the following functionalities: \\
    \quad \quad $\mathcal{F}_{\mathsf{Mul}}^{l_t, l_t}({s}_0, {c}_1)$ to learn $[\![ {sc} ]\!]^{2l_t}$.\\
    \quad \quad $\mathcal{F}_{\mathsf{Mul}}^{l_t, l_t}({c}_0, {s}_1)$ to learn $[\![ {cs} ]\!]^{2l_t}$.\\
    \quad \quad $\mathcal{F}_{\mathsf{Mul}}^{l_t, l_t}({c}_0, {c}_1)$ to learn $[\![ {cc} ]\!]^{2l_t}$.\\ 
    \quad \quad $\mathcal{F}_{\mathsf{Mul}}^{l_t, l_t}({s}_0, {s}_1)$ to learn $[\![ {ss} ]\!]^{2l_t}$.\\



    \For{$j \in \{0,1,2\}$}
    {
        $P_0$ and $P_1$ compute $C_j = \cos(-j \cdot \frac{L}{2^{f}})$ and $S_j = \sin(-j \cdot \frac{L}{2^{f}})$.

        $P_0$ and $P_1$ invoke $\mathcal{F}_{\mathsf{Mul}}^{2l_t, l_T}([\![ sc ]\!]^{2l_t} + [\![ cs ]\!]^{2l_t}, C_j)$ to learn $[\![ temp_0 ]\!]^{2l_t+l_T}$.
        
        $P_0$ and $P_1$ invoke $\mathcal{F}_{\mathsf{Mul}}^{2l_t, l_T}([\![ cc ]\!]^{2l_t} - [\![ ss ]\!]^{2l_t}, S_j)$ to learn $[\![ temp_1 ]\!]^{2l_t+l_T}$.

        $P_0$ and $P_1$ set $[\![ T ]\!]^{2l_t+l_T}[j] = ([\![ temp_0 ]\!]^{2l_t+l_T} + [\![ temp_1 ]\!]^{2l_t+l_T})$.

        $P_0$ and $P_1$ set $[\![ T ]\!]^{5 + f'}[j] = [\![ T ]\!]^{2l_t+l_T}[j] \gg (2f_t + f_T - f') \bmod 2^{5+f'}$.

    }

    $P_0$ and $P_1$ invoke $\mathcal{F}_{{\mathsf{MW}}}^{l,2}([\![ x ]\!]^{l})$ to learn $[\![ \mathsf{MW} ]\!]^{2}$.

    $P_0$ and $P_1$ invoke $\mathcal{F}_{{\mathsf{LUT}}}([\![ T ]\!]^{5+f'}, [\![ \mathsf{MW} ]\!]^{2})$ to learn $[\![ sin ]\!]^{5 + f'}$.



    $P_0$ and $P_1$ invoke $\mathcal{F}_{\mathsf{SExt}}^{5 + f', l'}([\![ sin ]\!]_i^{5 + f'})$ to learn $[\![ sin ]\!]_i^{l'}$.

    $P_0$ and $P_1$ output $[\![ sin ]\!]^{l'}$.

\end{algorithm}

\vspace{5pt}
\noindent
\textbf{Correctness and security.}
The correctness of $\Pi_{\mathsf{sin}}$ is ensured by Equation~\ref{equ:sin}.
The security comes from the security of protocols for 
$\mathcal{F}_{\mathsf{Mul}}$, 
$\mathcal{F}_{\mathsf{LUT}}$, 
$\mathcal{F}_{\mathsf{MW}}$, and 
$\mathcal{F}_{\mathsf{SExt}}$.

\vspace{5pt}
\noindent
\textbf{Complexity.}
$\Pi_{\mathsf{sin}}^{l, f, l', f'}$ invoke protocols for 
four $\mathcal{F}_{\mathsf{Mul}}^{l_t, l_t}$, 
two $\mathcal{F}_{\mathsf{Mul}}^{2l_t, l_T}$, 
one $\mathcal{F}_{\mathsf{MW}}^{l, 2}$, 
one $\mathcal{F}_{\mathsf{LUT}}$, and 
one $\mathcal{F}_{\mathsf{SExt}}^{3+f', l'}$.
The four $\mathcal{F}_{\mathsf{Mul}}^{l_t, l_t}$ can be realized by four $\mathcal{F}_{\mathsf{CrossTerm}}^{l_t, l_t}$, with communication $4 \cdot (l_t \cdot (\lambda + \frac{l_t}{2} + \frac{1}{2}) + l_t^2)$.
The two $\mathcal{F}_{\mathsf{Mul}}^{2l_t, l_T}$ can be realized by two $\mathcal{F}_{\mathsf{SExt}}^{2l_t, 2l_t + l_T}$, with communication $2(\lambda + l_T)$.
The communication for $\mathcal{F}_{\mathsf{MW}}$ is denoted as $C_{\mathsf{MW}}$.
Communication of protocols for $\mathcal{F}_{\mathsf{LUT}}$ and the last one $\mathcal{F}_{\mathsf{SExt}}^{5+f', l'}$ is $2\lambda + 4(2l_t + l_T)$ and $\lambda + l' - f - 3$.
Therefore, the total communication for $\Pi_{\mathsf{sin}}^{l, f, l', f'}$ is approximately $\lambda (4l_t + 5) + 6l_t^2 + 10 l_t + 6l_T + C_{\mathsf{MW}}$.

\section{Ideal Functionalities}


For our computing $\mathsf{MW}(x)$ protocol, the ideal functionality $\mathcal{F}_{\mathsf{MW}}$ is defined as follows.

\begin{tcolorbox}[
colback=white,
fonttitle=\bfseries,
title={Functionality $\mathcal{F}_{\mathsf{MW}}$},
coltitle=black,      
colbacktitle=white, 
]


\textbf{Constraint:} For $x \in \mathbb{Z}_L$, there is a known upper bound $B \leq \frac{L}{2}$, imposing $|x| < B$.
If there is no prior knowledge about the range of $x$, $B$ is set as $\frac{L}{2}$. 

\begin{itemize}
    \item 
    $\mathcal{F}_{\mathsf{MW}}$ receives $x_0$ from $P_0$ and $x_1$ from $P_1$.
    
    \item 
    $\mathcal{F}_{\mathsf{MW}}$ computes $mw = \mathsf{MW}(x_0, x_1, L)$ according to Equation~\ref{equ:MW def1}.

    \item 
    $\mathcal{F}_{\mathsf{MW}}$ shares $mw$ to $P_0$ and $P_1$, respectively.

\end{itemize}
\end{tcolorbox}

Based on $\mathsf{MW}(x)$ and $\mathsf{Real}(x)$, we can evaluate functions $\mathsf{func}(\cdot)$ with the property listed in Equation~\ref{equ:func x y z}.
The ideal functionality for evaluate $\mathcal{F}_{\mathsf{func}}$ is listed below, and the functionalities for evaluate integer division, trigonometric and exponential functions are instances of $\mathsf{func}(\cdot)$.

\begin{tcolorbox}[
colback=white,
fonttitle=\bfseries,
title={Functionality $\mathcal{F}_{\mathsf{func}}$},
coltitle=black,      
colbacktitle=white, 
]


\textbf{Constraint:} For $x \in \mathbb{Z}_L$, there is a known upper bound $B \leq \frac{L}{2}$, imposing $|x| < B$.
If there is no prior knowledge about the range of $x$, $B$ is set as $\frac{L}{2}$.

\begin{itemize}
    \item 
    $\mathcal{F}_{\mathsf{func}}$ receives $x_0$ from $P_0$ and $x_1$ from $P_1$.

    \item 
    $\mathcal{F}_{\mathsf{func}}$ computes $mw = \mathsf{MW}(x_0, x_1, L)$ according to Equation~\ref{equ:MW def1} and $res=\sum_{i=0}^{k-1} f_i(\frac{x_0}{2^f}) \cdot g_i(\frac{x_1}{2^f}) \cdot h_i(\frac{-\mathsf{MW}(x) \cdot L}{2^f}).$

    \item 
    $\mathcal{F}_{\mathsf{func}}$ shares $res$ to $P_0$ and $P_1$, respectively.

\end{itemize}

\end{tcolorbox}
